\documentstyle[12pt,aaspp]{article}

\begin{document}

%
%
\def\prd#1#2#3#4{#4 19#3 Phys.~Rev.~D,\/ #1, #2 }
\def\pl#1#2#3#4{#4 19#3 Phys.~Lett.,\/ #1, #2 }
\def\prl#1#2#3#4{#4 19#3 Phys.~Rev.~Lett.,\/ #1, #2 }
\def\pr#1#2#3#4{#4 19#3 Phys.~Rev.,\/ #1, #2 }
\def\prep#1#2#3#4{#4 19#3 Phys.~Rep.,\/ #1, #2 }
\def\pfl#1#2#3#4{#4 19#3 Phys.~Fluids,\/ #1, #2 }
\def\pps#1#2#3#4{#4 19#3 Proc.~Phys.~Soc.,\/ #1, #2 }
\def\nucl#1#2#3#4{#4 19#3 Nucl.~Phys.,\/ #1, #2 }
\def\mpl#1#2#3#4{#4 19#3 Mod.~Phys.~Lett.,\/ #1, #2 }
\def\apj#1#2#3#4{#4 19#3 Ap.~J.,\/ #1, #2 }
\def\aj#1#2#3#4{#4 19#3 Astr.~J.,\/ #1, #2}
\def\acta#1#2#3#4{#4 19#3 Acta ~Astr.,\/ #1, #2}
\def\rev#1#2#3#4{#4 19#3 Rev.~Mod.~Phys.,\/ #1, #2 }
\def\nuovo#1#2#3#4{#4 19#3 Nuovo~Cimento~C,\/ #1, #2 }
\def\jetp#1#2#3#4{#4 19#3 Sov.~Phys.~JETP,\/ #1, #2 }
\def\sovast#1#2#3#4{#4 19#3 Sov.~Ast.~AJ,\/ #1, #2 }
\def\pasj#1#2#3#4{#4 19#3 Pub.~Ast.~Soc.~Japan,\/ #1, #2 }
\def\pasp#1#2#3#4{#4 19#3 Pub.~Ast.~Soc.~Pacific,\/ #1, #2 }
\def\annphy#1#2#3#4{#4 19#3 Ann. Phys. (NY), \/ #1, #2 }
\def\yad#1#2#3#4{#4 19#3 Yad. Fiz.,\/ #1, #2 }
\def\sjnp#1#2#3#4{#4 19#3 Sov. J. Nucl. Phys.,\/ #1, #2 }
\def\astap#1#2#3#4{#4 19#3 Ast. Ap.,\/ #1, #2 }
\def\anrevaa#1#2#3#4{#4 19#3 Ann. Rev. Astr. Ap.,\/ #1, #2 }
\def\mnras#1#2#3#4{#4 19#3 M.N.R.A.S.,\/ #1, #2 }
\def\jdphysics#1#2#3#4{#4 19#3 J. de Physique,\/ #1,#2 }
\def\jqsrt#1#2#3#4{#4 19#3 J. Quant. Spec. Rad. Transfer,\/ #1,#2 }
\def\jetpl#1#2#3#4{#4 19#3 J.E.T.P. Lett.,\/ #1,#2 }
\def\apjl#1#2#3#4{#4 19#3 Ap. J. (Letters).,\/ #1,#2 }
\def\apjs#1#2#3#4{#4 19#3 Ap. J. (Supp.).,\/ #1,#2 }
\def\apl#1#2#3#4{#4 19#3 Ap. Lett.,\/ #1,#2 }
\def\astss#1#2#3#4{#4 19#3 Ap. Sp. Sci.,\/ #1,#2 }
\def\nature#1#2#3#4{#4 19#3 Nature,\/ #1,#2 }
\def\spscirev#1#2#3#4{#4 19#3 Sp. Sci. Rev.,\/ #1,#2 }
\def\advspres#1#2#3#4{#4 19#3 Adv. Sp. Res.,\/ #1,#2 }
%
%
%
\def\Msun{M_{\odot}}
\def\Mdot{\dot M}
\def\deg{$^\circ$\ }
\def\etal{{\it et~al.\ }}
\def\eg{{\it e.g.,\ }}
\def\etc{{\it etc.}}
\def\ie{{\it i.e.,}\ }
\def\ksec{{km~s$^{-1}$}}
\def\arcsec{{$^{\prime\prime}$}}
\def\arcmin{{$^{\prime}$}}
\def\subsun{_{\twelvesy\odot}}
\def\sun{\twelvesy\odot}
\def\gtwid{\mathrel{\raise.3ex\hbox{$>$\kern-.75em\lower1ex\hbox{$\sim$}}}}
\def\ltwid{\mathrel{\raise.3ex\hbox{$<$\kern-.75em\lower1ex\hbox{$\sim$}}}}
\def\plusminus{\mathrel{\raise.3ex\hbox{$+$\kern-.75em\lower1ex\hbox{$-$}}}}
\def\minusplus{\mathrel{\raise.3ex\hbox{$-$\kern-.75em\lower1ex\hbox{$+$}}}}
%
%
%
%
\title{The Vertical Structure and Ultraviolet Spectrum of X-ray Irradiated 
Accretion Disks in Active Galactic Nuclei}

\author{Mark W. Sincell}
\affil{Department of Physics MC 704 \\The University of Illinois at
Urbana-Champaign \\1110 W. Green Street \\Urbana, IL 61801-3080}
\authoraddr{Department of Physics MC 704 \\The University of Illinois at
Urbana-Champaign \\1110 W. Green Street \\Urbana, IL 61801-3080}

\author{Julian H. Krolik}
\affil{Department of Physics \& Astronomy \\The Johns Hopkins University 
\\Baltimore, MD 21218}
\authoraddr{Department of Physics \& Astronomy \\The Johns Hopkins University 
\\Baltimore, MD 21218}

\begin{abstract}

Motivated by recent work indicating that the UV continuum in AGN may be
produced by reradiation of energy absorbed from X-rays irradiating
an accretion disk, we present a calculation of the vertical
structures and ultraviolet spectra
of X-ray irradiated accretion disks around massive non-rotating black holes.
After finding the radial dependence
of vertically-integrated quantities for these disks, we solve the equations of
hydrostatic equilibrium, energy balance, and frequency-dependent
radiation transfer as functions of altitude.  To solve the last set of
equations, we use a variable Eddington factor method.  We include
electron scattering, free-free, and HI, HeI, and HeII bound-free opacities and
the corresponding continuum cooling processes.

The incident X-ray flux heats a thin
layer of material 3-4 scale heights above the midplane of the disk.
This X-ray heated skin has two layers: a radiation pressure supported region
in which the UV flux is created, and, immediately above this layer,
a warmer zone, optically thin to UV radiation, formed where the X-ray 
ionization parameter is large.  In the lower layer the gas pressure
is nearly independent of altitude but the temperature increases upward. 

The fraction of the incident hard X-ray flux which emerges in the UV falls
with increasing $\dot m$ (the accretion rate in Eddington units).
At frequencies below the
Lyman edge the slope of the continuum ($d\ln L_\nu/d\ln\nu$) varies from
-1.6 to 0.8 as $\dot m/m_8$ increases from 0.001 to 1.
Here $m_8$ is the mass of the central black hole in units of $10^8 M_{\odot}$.
In all cases examined ($0.003 \leq \dot m \leq 0.3$ and $0.27 \leq m_8
\leq 27$), the Lyman edge appears in emission.  
The amplitude of the Lyman edge feature increases with $m_8$
but is relatively independent of $\dot m$.
The amplitude of the Lyman edge emission feature increases
with disk inclination.
Compton scattering in disk coronae can smooth the Lyman edge 
feature only if $\tau_c \gtwid 0.5$, where $\tau_c$ is the Thomson depth 
of the coronae.

   While the overall spectral shape predicted by X-ray irradiation may be
compatible with observations, the Lyman edge emission feature it predicts
is not. 
This finding raises questions for many otherwise plausible models
in which X-ray irradiation plays a major role.

\end{abstract}

\section{Introduction}

The optical and UV emission  of radio-quiet active galactic nuclei (AGNs)
is dominated by a
quasi-thermal component, the ``Big Blue Bump" (Shields 1978, Malkan \& Sargent
1982, Malkan 1983).  The Big Blue Bump is usually interpreted as 
thermal emission from a geometrically
thin, optically thick, accretion disk around a massive black hole
(Malkan 1983, Sun \& Malkan 1989, Laor \& Netzer 1989: LN89).
This model makes two key predictions.  
First, the large jump in the absorptive opacity at the Lyman edge combined
with the vertical temperature gradients in the disk insure that some feature
will be present at the Lyman edge (Kolykhalov \& Sunyaev 1984,
LN89).  For the same reason, some have predicted that the polarization of the
disk should decrease at frequencies above the Lyman edge
(\eg Laor, Netzer \& Piran 1990; but see Blaes \& Agol 1996 for another view).
Second, variations in different portions
of the optical/ultraviolet band can be associated with fluctuations in
the conditions at particular radii.  If fluctuations move radially
through the disk, there should be time-delays between variations seen
at different wavelengths which correspond to signal travel times within
the disk.

  Unfortunately, while this picture is very attractive, it suffers from
severe problems, both theoretical and observational.  From the theoretical
point of view, it is generally expected that the viscous stress is
proportional to the pressure (the ``$\alpha$-model" introduced
by Shakura \& Sunyaev 1973: SS73), including
the radiation pressure when the disk is optically thick.  Unfortunately,
if this is true, when the accretion rate is more than a small fraction of
the Eddington rate, the disk is thermally unstable throughout its inner region
(Shakura \& Sunyaev 1976).  For this reason, the most detailed disk spectral
calculations made to date (LN89) supposed that the stress was instead
proportional to the geometric mean of the gas and radiation pressures.

In addition, neither of the two key predictions just described is
confirmed by observations.
Partial Lyman edges have been detected in very few AGN (Antonucci, Kinney
\& Ford 1989; Koratkar, Kinney \& Bohlin 1992) and the polarization of two
quasars with partial absorption edges {\it rises} across the Lyman edge
(Koratkar, \etal 1995). 
Intensive monitoring of the type 1 Seyfert galaxy NGC 5548 revealed that
variations from the optical to the far-UV are tightly correlated.  The
signal travel speed through a conventional disk that would be required
to coordinate these variations was found to be at least $\sim 0.1c$,
far higher than any of the expected signal speeds (Krolik \etal 1991).
Moreover, variations in the soft X-ray flux have been found to be almost
as tightly correlated with ultraviolet fluctuations (Clavel, \etal 1992).  
These two findings suggest that the Big Blue Bump may actually be due
to reprocessed X-ray emission (\eg Rokaki, Collin-Souffrin, \& Magnan 1993).

Other evidence also points to the importance of X-ray reprocessing in the
formation of AGN continua.  In many type 1 Seyfert galaxies, there is
a spectral feature in the vicinity of 10 -- 30 keV which is readily interpreted
as ``Compton reflection" of X-rays from a cool, optically thick surface
occupying roughly half of the solid angle around the X-ray source (Pounds
\etal 1990).  It would be very natural to interpret this reflector as an
accretion disk (Lightman \& White 1988) (but a number of other interpretations
are also possible: Nandra \& George 1994; Krolik, Madau, \& ${\rm\dot Zycki}$ 
1994; Ghisellini, Haardt \& Matt 1994).  Further support for the thought
that the disk is subjected to intense X-ray irradiation comes from the
discovery that the Fe K$\alpha$ profiles in many AGN are extremely broad
(\eg as summarized by Nandra 1997).

Recently popular models for the production of the X-rays are also
consistent with this picture in which a large fraction of the emitted X-rays
strike a nearby cool surface, possibly the accretion disk (Haardt \& Maraschi 
1993: HM93; Haardt, Maraschi \& Ghisellini 1994; Pietrini \& Krolik 1995; Stern
\etal 1995).  These models suppose that a large part of the disk's dissipation
takes place in a small amount of mass at or above the disk's surface.  The
very large heating rate per unit mass in an optically thin gas creates
a very hot corona which cools by inverse Compton scattering.  Roughly half
the X-ray flux generated strikes the disk, and is reprocessed into UV
photons which then return to the corona as seeds for inverse Compton scattering.
Fits to the broad-band X-ray spectra of AGN are at least qualitatively
consistent with this sort of thermal Comptonization model (e.g. Zdziarski
\etal 1995).

In its simplest form, the corona model 
predicts that the observed UV flux should be roughly equal to the observed
intrinsic X-ray flux (i.e. the X-ray flux after subtracting any components
due to reprocessing).  Whether this prediction is consistent with
the observations is uncertain.  
The UV/X-ray flux ratio is conventionally
parameterized by $\alpha_{ox}$, the spectral index of a power-law
interpolated between 2500\AA{} and 2 keV, as measured in the AGN rest-frame.
Depending on the sample definition, the observed distribution of $\alpha_{ox}$
varies somewhat, but typically it ranges from $\simeq 1.2$ to $\simeq 2.1$,
with a tendency for lower luminosity AGN to have smaller $\alpha_{ox}$
(Stocke et al. 1991; Wilkes et al. 1994; Green et al. 1995).
The flux at at 2500\AA{} is $\simeq 20 \times 400^{\alpha_{ox} - 1.5}$
times the flux at 2 keV so that, if we compare at these two frequencies alone,
we would conclude that the UV luminosity exceeds the 
X-ray luminosity in most AGN.  However, to find the ratio of the total UV
and X-ray fluxes, substantial---and very uncertain---bolometric corrections
must be applied to both.  We cannot observe low-redshift
AGN in the rest-frame EUV, but composite high-redshift quasar
spectra (e.g. Zheng et al. 1996) suggest that the spectrum rolls over sharply
at wavelengths shorter than $\simeq 1000$~\AA.  If this shape is general, the UV
bolometric correction (relative to $\nu F_\nu$ at 2500\AA)
might be as small as $\simeq 2$, but there is likely to be a considerable
dispersion from case to case.  The X-ray bolometric correction
is similarly uncertain.  Those low luminosity AGN for which OSSE has
been able to obtain spectra typically show power-laws with indices
$\simeq 0.9$ across the hard X-ray range,
supplemented by a substantial bump at a few tens of keV which is generally
believed to be due to reprocessing (Zdziarski et al. 1995).
If that power-law extends from 0.5 to 200 keV,
the bolometric correction relative to $\nu F_\nu$ at 2 keV is 6.5.
Taking these bolometric corrections at face value, we would conclude
that the UV and X-ray fluxes in low luminosity AGN (where the evidence for
the corona model is best) are indeed close to being equal,
while the UV flux is typically rather greater
than the intrinsic X-ray flux in higher luminosity AGN.  However, it is
clear from the character of this estimate that it is very weakly based. 
The situation is
made still cloudier by the possibility that relativistic effects may
direct the majority of the emitted X-rays toward the disk, reducing
the observed ratio between X-rays and UV (Martocchia and Matt 1996).

   It is our goal to begin the process of testing these coronal models by
comparing the ultraviolet spectrum they predict with observations.  In
this paper we develop the techniques necessary to compute the predicted
UV spectra, and apply them to the simplest version of the model: that in which
{\it all} the dissipation takes place in a slab-like
corona resting directly on top of the accretion disk.  In future work
we will examine other variations, such as models in which the dissipation is
shared between the disk proper and the corona, and models with more
complicated coronal geometry. 

  An immediate consequence of this picture in which most of the dissipation
takes place in a corona, and not inside the disk proper, is that
radiation pressure support in the bulk of the disk is far weaker
than in a conventional disk.  The radiation force is simply the product of the
radiation flux and the opacity; where there is no outgoing flux (because
there is no internal heat generation), there can be no radiation force.
As a result, such disks collapse to a state of much greater density
(and also column density) than would be predicted by conventional models
(Svensson \& Zdziarski 1995: SZ95).  They are then supported primarily by gas
pressure, and (in the $\alpha$-model) are thermally stable.
It is also an immediate corollary that
these disks are very nearly isothermal---zero outward flux in an optically
thick environment implies constant temperature.

  Some analogous work has been done in the past.  A number of groups (e.g.
Ross \& Fabian 1993; ${\rm \dot Zycki}$ \etal 1994) have computed the spectra of
X-ray irradiated disks, but all have assumed the disks are radiation pressure
supported.  Because disks in which most dissipation is in a corona
are much denser than conventional radiation pressure-supported disks,
the character of the UV spectrum they produce is quite different from what
has been predicted by assuming a conventional disk structure.  Others
have tried to predict the UV spectra radiated by dense lumps heated 
by X-rays (Guilbert
\& Rees 1988, Ferland \& Rees 1988).

In \S\ref{sec: isothermal disk} we solve the structure equations
for X-ray irradiated disks.  The solution of the radiative transfer and vertical
structure equations in the X-ray heated skin is described in \S
\ref{sec: xray skin} and our results are summarized in \S\ref{sec: results}
The gravitational stability of the cold disk is discussed in 
\S\ref{sec: gravitational instability}
We conclude in \S\ref{sec: conclusions}  

\section{The Structure of X-ray Irradiated Accretion Disks}
\label{sec: isothermal disk}

When there is no internal dissipation (SZ95
treat the case in which the dissipation is shared between
the disk and the corona), the main body of the disk is rigorously
isothermal, as explained in \S 1.  The radial structure of geometrically thin 
isothermal accretion disks
is defined by two equations: the conservation of angular momentum
\begin{equation}
W_{r \phi} = {\dot M \omega \over 2\pi} 
{Q_{NT} C_{NT}^{1/2} \over B_{NT} D_{NT} },
\end{equation}
and the inter-ring stress prescription, for which we use the $\alpha$-model: 
\begin{equation}
W_{r \phi} = \alpha_{SS} \Sigma_o c_s^2 .
\end{equation}
Vertical structure is determined by hydrostatic equilibrium:
\begin{equation}
{\partial P \over \partial z} = - {G M \rho(z) \over R^2} \left( {z \over R} 
\right) \left( {C_{RH} \over B_{RH}} \right)
\end{equation}
and the isothermal equation of state.
The symbols in these equations represent the mass of the central black hole 
($M$), the mass accretion rate ($\dot M$), the Keplerian angular velocity 
($\omega$), the vertically integrated shear stress ($W_{r\phi}$),
the radial coordinate ($R$), the vertical coordinate ($z$), 
the gas density ($\rho$), the gas
pressure ($P$), the sound speed in the gas ($c_s$), the surface mass density
\begin{equation}
\Sigma_o = \int_0^{\infty} \rho(z) dz,
\end{equation}
and the dimensionless stress $\alpha_{SS}$.
The general relativistic correction factors are  taken from the standard
references.  The
subscript NT refers to the Novikov and Thorne (1973, Page \& Thorne 1974) 
expressions and the 
subscript RH refers to the recent corrections presented by Rifferts \& Herrold
(1995).

These equations are easily solved.  
The integrated surface mass density is
\begin{equation}
\label{eq: sigma_o}
\Sigma_{o} = { \dot M \omega \over 2\pi \alpha_{SS} c_s^2} 
{Q_{NT} C_{NT}^{1/2} \over B_{NT} D_{NT} }
\end{equation}
and the density profile is Gaussian
\begin{equation}
\rho(z) = \rho_{o} e^{-(z/z_o)^2}
\end{equation}
with a scale height of 
\begin{equation}
\label{eq: disk scale height}
z_o^2 = {2kT_g R^3 \over G M m_p} \left( { B_{RH} \over C_{RH} } \right)
\end{equation}
where $T_g$ is the gas temperature in the accretion disk.
The normalization of the density,
\begin{equation}
\label{eq: density normalization}
\rho_o = {2 \over \sqrt{\pi} } {\Sigma_o \over z_o},
\end{equation}
is set by assuming that all the mass lies in the isothermal inner disk.  

It is instructive to express the disk solution in terms of nondimensional
parameters
\begin{equation}
\label{eq: disk parameter scalings}
\dot m = {\dot M \over \dot M_E} \qquad m_8 = {M \over 10^8 \Msun} \qquad
r = {R \over R_s} \qquad T_5 = {T_g \over 10^5 K}
\end{equation}
where we have introduced the Eddington accretion rate
\begin{equation}
\dot M_E = {4 \pi c R_s \over \eta \kappa_{es}} = 2.44 \times 10^{26} \, m_8
\quad {\rm g/s}
\end{equation}
and the Schwarzschild radius
\begin{equation}
R_s = {2GM \over c^2} = 2.95 \times 10^{13} \, m_8 \quad {\rm cm}.
\end{equation}
The electron scattering opacity is $\kappa_{es}$.
The numerical factors have been calculated by assuming an accretion efficiency
of $\eta = 0.0572$, which is appropriate for a non-rotating black hole.
Using these scalings we find that
\begin{equation}
\label{eq: sigma0}
\Sigma_o = 3.38 \times 10^9 \, \dot m r^{-3/2} \alpha_{SS}^{-1} T_5^{-1}
\left( {Q_{NT} C_{NT}^{1/2} \over B_{NT} D_{NT} } \right) \quad
{\rm gm/cm^2},
\end{equation}
\begin{equation}
z_o = 5.65 \times 10^9 \, T_5^{1/2} m_8 r^{3/2}
\left( {C_{RH} \over B_{RH} } \right)^{-1/2} \quad
{\rm cm},
\end{equation}
and
\begin{equation}
\label{eq: isothermal gas density}
\rho_o = 0.68 \, \dot m m_8^{-1} r^{-3} \alpha_{SS}^{-1} T_5^{-3/2}
\left( {Q_{NT} C_{NT}^{1/2} \over B_{NT} D_{NT} } \right)
\left( {C_{RH} \over B_{RH} } \right)^{1/2} \quad
{\rm gm/cm^3}.
\end{equation}
There are two important things to note about these results.  First, the
total surface mass density is many orders of magnitude larger than the
penetration depth of the hard X-rays.  Consequently, supposing the bulk
of the disk to be isothermal and gas pressure-supported is an excellent
approximation.  Second, the gas density of the isothermal disk
is very much larger than the $\sim 10^{-10}$~gm cm$^{-3}$ typical of
conventional radiation pressure supported disks (\eg SS73).

\section{The X-ray Heated Skin}
\label{sec: xray skin}

     Few X-rays can penetrate much greater than the Compton depth
$\Sigma_C \simeq 0.5$~gm~cm$^{-2}$.   On the other hand, as shown by
equation 12, these disks are extremely optically thick to Compton scattering. 
Consequently, there are essentially no X-rays below $z/z_o \simeq [\ln(\Sigma_o/
\Sigma_C)]^{1/2} \simeq 3$ -- $4$.  The exact number of scale
heights above the midplane where the X-ray-heated skin begins depends
only slightly on parameters.  In this estimate, and all the numerical work,
we set $\alpha_{SS} = 0.1$ for definiteness, but it hardly affects the results 
at all.

\subsection{Differential Equations and Constraints}
  Inside the X-ray-heated skin, the temperature, pressure, and radiation
intensity are determined as functions of altitude by solving the 
differential equations for radiative transfer,
\begin{equation}
\label{eq: radiative transfer}
{\partial \over \partial \Sigma} \left\{ {1 \over 
\kappa_{T,\nu}} {\partial (f_{\nu} J_{\nu})
\over \partial \Sigma} \right\} = \kappa_{a,\nu} J_{\nu} - {1 \over 4 \pi}
\varepsilon_{\nu}  
\end{equation}
hydrostatic equilibrium
\begin{equation}
\label{eq: hydrostatic equilibrium}
{\partial P \over \partial \Sigma} = g(\Sigma) + {H(\Sigma) \over 3c}
- {4\pi \over c} \int d\nu {\partial (f_{\nu} J_{\nu}) \over \partial \Sigma}
\end{equation}
and the surface mass density
\begin{equation} 
\label{eq: surface mass density}
{\partial z \over \partial \Sigma} = - {1 \over \rho} 
\end{equation}
subject to the local constraints of radiative equilibrium
\begin{equation}
\label{eq: radiative equilibrium}
\int d\nu \left\{4\pi \kappa_{a,\nu} J_{\nu} - \varepsilon_{\nu} \right\}
+ H(\Sigma) = 0
\end{equation}
and charge conservation
\begin{equation}
\label{eq: charge conservation}
n_e = \sum_k \sum_{l=1}^{L_k} l n_{lk}.
\end{equation}
Equations \ref{eq: radiative transfer}, \ref{eq: hydrostatic equilibrium},
\ref{eq: surface mass density}, \ref{eq: radiative equilibrium}, 
\ref{eq: charge conservation} are solved for 
the mean ultraviolet intensity ($J_{\nu}$), gas pressure ($P$),
gas temperature ($T$), electron density ($n_e$) and vertical
coordinate ($z$) as functions of the surface mass density.  
The other symbols represent the X-ray heating rate ($H(\Sigma)$,
see \S \ref{subsec: heating rate}),
total opacity ($\kappa_{T,\nu}$),
the absorptive opacity ($\kappa_{a,\nu}$), the emissivity ($\varepsilon_{\nu}$),
the variable Eddington factors 
($f_{\nu}$), the number density of ion species $k$ in ionization state $l$ 
($n_{lk}$)
and the maximum ionization state of species $k$ ($L_k$). 

We assume that the disk is geometrically thin and neglect self-gravity of the
disk.
Under these assumptions, the gravitational acceleration of the material in 
the disk is (Rifferts \& Herrold 1995)
\begin{equation}
g(\Sigma) = {G M z(\Sigma) \over R^3} \left( {C_{RH} \over B_{RH}} \right).
\end{equation}
The gravitational force of the underlying isothermal disk can contribute to
the net acceleration of material in the skin but the contribution is
small (see the discussion in 
\S\ref{sec: isothermal disk}).  
The momentum of the 
absorbed X-rays is transferred to the gas, resulting in an additional force
directed towards the disk midplane.  
The gradient of the hard X-ray radiation pressure is
\begin{equation}
{\partial P_{r,x} \over \partial \Sigma} = - {H(\Sigma) \over 3c}
\end{equation}
in the Eddington approximation, which we have used to solve for $H(\Sigma)$
(\S\ref{subsec: heating rate}).

The total opacity includes bremsstrahlung, photoionization and electron 
scattering.  We neglect all line opacities.  The expression for the 
free-free opacity is taken from Rybicki \& Lightman (1979) and the
general expression for the photo-ionization opacity is from Mihalas (1978).
The bound-free oscillator strengths of HI and HeII are calculated using the
analytic formulae for hydrogen-like atoms (\eg Mihalas 1978) and the HeI
oscillator strengths are from Bell \& Kingston (1967).  
We calculate the number densities
of electrons, HI, HII, HeI, HeII and HeIII assuming
Saha equilibrium, \ie local thermodynamic equilibrium,
and hydrogen and helium abundances of X=0.9 and Y=0.1,  respectively.  Explicit
photoionization and recombination are not included in the  calculation of the
ionization state.

We compute the emissivity using the corresponding continuum cooling processes:
bremsstrahlung and HI, HeI, and HeII radiative recombination.  
The gas density in the X-ray heated layer is
high enough that we expect resonance line cooling to be
suppressed.
Resonance lines are thermalized 
when the collisional deexcitation rate is greater than the product of the
radiation rate and the photon escape probability.  
Neglecting stimulated
emission, this occurs when the fractional abundance of the lower state
$X_l$ satisfies the inequality
\begin{equation}
X_l > 10^{-9} n_{e18}^{-1} N_{H22} \left( {m_A \over m_p} \right)^{-1/2} 
(h\nu/I_H)^3 , 
\end{equation}
where $n_{e18}$ is the electron density in units of $10^{18}$ cm$^{-3}$,
$N_{H22}$ is the column density of H atoms in units of $10^{22}$ cm$^{-2}$, and
$m_{A,p}$ the masses of the atom and proton, respectively.  
These scalings were chosen because the gas density at 3.5 scale heights 
from the midplane is typically
$n_e \sim 10^{18}$ cm$^{-3}$ 
and 2 keV photons are absorbed at a column density $N \sim 10^{22}$ cm$^{-2}$.
When the temperature is $\sim 10^5$~K UV and EUV lines
are the important potential coolants, but the only lines which contribute 
to the cooling
are those with very low abundance lower states.

By contrast, the
density in the atmosphere of an irradiated radiation pressure supported
disk is characteristically four orders of magnitude lower.
This means that the ionization parameter is so much greater that a thermal
runaway is triggered, and soft X-ray lines are the most important transitions
for cooling.  Due to the combination of lower density and higher radiation
rates, emission lines are much more important in the thermal balance in this
context than in our limit of the zero internal dissipation disk.

In principle, the complete angular dependent radiative transport problem can
be solved by introducing variable Eddington factors (\eg Mihalas 1978).  The 
computational overhead associated with an iterative calculation of 
the
Eddington factors is fairly large and, given the other approximations we have
made, probably unwarranted.
Instead, we adopt a simple parametrization
\begin{equation}
\label{eq: analytic fedd}
f_{\nu} = { {1 + \tau_{\nu}^2} \over {1 + 3 \tau_{\nu}^2} },
\end{equation}
where
\begin{equation}
\tau_{\nu} = \int \kappa_{T,\nu} d\Sigma,
\end{equation}
which has the correct limiting behavior (G. S. Miller, private communication).
We perform an iterative calculation of $f_{\nu}$ 
for one set of disk parameters in
\S\ref{subsubsec: angular distribution}
and use this solution to test the accuracy of the analytic prescription.

The complete system of equations is solved by transforming the differential
equations into finite-difference equations, linearizing with respect to all
variables, and relaxing from an initial solution.  
We solve the structure 
equations  for a grey atmosphere with a shooting procedure (Press,
\etal 1992) and use this solution as a first guess for the relaxation method.
The only significant difference between our work and the standard stellar
atmosphere calculation 
(\eg Mihalas 1978, Rybicki 1971)
is the non-zero heating rate (see eq. 
\ref{eq: radiative equilibrium}).

\subsection{The X-ray Heating Rate}
\label{subsec: heating rate}
We assume that all the energy dissipated due to accretion at radius $R$
is released as heat in the corona.  The rate per unit area is
\begin{equation}
\label{eq: dissipation rate}
Q_{dis} = {3GM \dot M \over 8\pi R^3} 
\left( {Q_{NT} \over B_{NT} C_{NT}^{1/2} } \right),
\end{equation}
and this is converted into an X-ray flux 
by inverse Compton scattering of the UV photons emerging from the disk (HM93).
One-half of this flux illuminates the underlying disk.  
In terms of the dimensionless parameters defined in eq. 
\ref{eq: disk parameter scalings}
\begin{equation}
\label{eq: scaled dissipation rate}
Q_{dis} = 1.50 \times 10^{19} \, \dot m m_8^{-1} r^{-3} 
\left( {Q_{NT} \over B_{NT} C_{NT}^{1/2} } \right)
\mbox{ergs ${\rm cm^{-2} s^{-1}}$}.
\end{equation}

    The spectrum of the illuminating flux is taken to be
\begin{equation}
S(E) = S_o E^{-\alpha} \exp{-E/E_c}  
\end{equation}
where $E_c=10 {\rm keV}$.   A self-consistent calculation of the X-ray
spectrum is beyond the scope of this paper, so we simply assume
$\alpha = 0.9$, as is commonly observed (Zdziarski \etal 1995).
The exponential cut-off is meant to mimic the sharp rise in the X-ray albedo
above 10~keV (Lightman \& White 1988).  As a result of the high albedo
for harder photons, they are mostly reflected and do not contribute to
the heating rate.  Still harder photons ($E > 100$~keV) lose much of
their energy in recoil, but because there are many uncertainties
in our knowledge of the high energy spectrum in AGN (Zdziarski \etal 1995),
and because the expected contribution to the net heating is small and 
depends sensitively on the high energy spectrum
(P. ${\rm Z\dot ycki}$, private communication),
we neglect this contribution. We also suppose that
there is no sub-structure to the corona; the local heating rate could vary
substantially if the dissipation is concentrated
in active regions (\eg Haardt, Maraschi \& Ghisellini 1994).
These complications will be addressed in later work.

Given this simple parametrization, the heating rate is a well-defined function 
of the surface mass density
\begin{equation}
\label{eq: local heating rate}
H(\Sigma) = \int \kappa_{x,eff}(E) S(E) e^{-\kappa_{x,eff}(E)
\Sigma} dE
\end{equation}
where the effective X-ray absorption opacity is defined as (Rybicki \& 
Lightman 1979)
\begin{equation}
\kappa_{x,eff}(E) = \left[ 3 \kappa_x (E) \left\{ \kappa_x (E) +
\kappa_{es} \right\} \right] ^{1/2}.
\end{equation}
and $\kappa_{es}$ is the electron scattering opacity.  

The X-ray
opacity in this energy range is dominated by photoionization of oxygen and
iron (\eg London, McCray \& Auer 1981).  The threshold energy for K-shell 
photoionization depends upon the ionization state of the gas (Daltabuit
\& Cox 1972) which is largely determined by the incident X-ray flux 
(Lightman \& White 1988, ${\rm Z\dot ycki}$, \etal 1994).  
A careful calculation of the ionization balance
of the heavy elements is beyond the scope of this paper.  Instead, we assume
threshold energies of 0.8 keV and 8.0 keV for oxygen and iron, respectively,
and solar abundances of these elements.  The X-ray opacity can
then be estimated as
\begin{equation}
\kappa_x (E) = \left\{ \begin{array}{ll}
  47.9 \left( {0.8 {\rm keV} \over E} \right)^3 \mbox{${\rm cm^{2} g^{-1}}$}
           & \mbox{$0.8 < E < 8.0$ keV}\\
  0.17 \left( {8.0 {\rm keV} \over E} \right)^3 \mbox{${\rm cm^{2} g^{-1}}$}
           & \mbox{$E > 8.0$ keV}
      \end{array} \right.
\end{equation}
using absorption cross-sections for hydrogen-like atoms
(Rybicki \& Lightman 1979).  This approximate description slightly
underestimates the absorptive opacity in most of the X-ray-heated atmosphere,
where most medium-Z atoms retain some valence shell electrons.

The expression for $H(\Sigma)$ (eq. \ref{eq: local heating rate}) 
is equivalent to a two-stream solution for the radiative transfer equation in
the Eddington approximation.  This approximation is reasonable because the
incident hard X-rays are nearly isotropic in the lower hemisphere and electron
scattering in the disk will tend to isotropize the X-rays.  The heating rate
at lower $\Sigma$ is probably underestimated in this approximation because
X-rays incident at large angles will be absorbed at smaller $\Sigma$.

The local heating rate is calculated by integrating equation 
\ref{eq: local heating rate} numerically and the results for the assumed
X-ray spectrum are plotted in fig \ref{fig: xray heating rate}.  The overall
scale of the heating rate depends upon the accretion parameters, but the 
$\Sigma$ dependence is the same for all the models.  At low column densities, 
the heating rate is almost independent of depth because all X-ray
frequencies are optically thin.  When $0.01~$gm~cm$^{-2}
 \ltwid \Sigma \ltwid 1.0$~gm~cm$^{-2}$,
\begin{equation}
H(\Sigma) \propto E_* \kappa_{x,eff}(E_*) S(E_*) \propto
 \Sigma^{-(2 + \alpha)/3} \simeq \Sigma^{-1}
\end{equation}
because photons of energy $E_*$ are absorbed at the depth
$\Sigma(E_*) \propto E_*^3$.
At $\Sigma \gg 1.0$~gm~cm$^{-2}$, the X-ray flux is exponentially
attenuated and the heating rate falls rapidly.  The analytic approximation
\begin{equation}
\label{eq: analytic heating rate}
H(\Sigma) \simeq 3.3 Q_{dis} \left\{ 1 - \exp(-0.025/\Sigma) \right\}
\exp(-\Sigma/11.0)
\end{equation}
is also plotted in fig. \ref{fig: xray heating rate}.  Here $\Sigma$ is
measured in units of gm~cm$^{-2}$.

\subsection{Boundary Conditions}
The boundary between the X-ray heated skin and the isothermal disk is
set at $\Sigma_1 = 50.0 \mbox{gm ${\rm cm^{-2}}$}$.  
The X-ray heating rate at the boundary is negligible 
($\Sigma_1 \gg 1 \mbox{gm ${\rm cm^{-2}}$}$) and we assume no dissipation
in the cold disk, so 
\begin{equation}
F_{\nu} = { 4\pi \over \kappa_{T,\nu} } {\partial (f_{\nu} J_{\nu})
 \over \partial \Sigma} = 0
\end{equation}
at the inner boundary. 
The gas temperature is taken to be continuous across the boundary and
the inner boundary conditions on the gas pressure and vertical coordinate 
are fixed by requiring that the numerical solution for the skin
match the analytic solution for the isothermal disk at $\Sigma_1$.  

The X-ray heated skin has a sharp outer boundary because a thermal
runaway occurs at that point (see \S 4.1 for a discussion of its physical
origin).  We designate the column density at this runaway $\Sigma_2$.
Its location is found by bisection.  We guess $\Sigma_2$ and
attempt a solution.  If the code converges to a thermally stable solution, 
$\Sigma_2$ is decreased; if the code fails to converge, 
it is increased.  This is repeated until the fractional change in 
$\Sigma_2$ is less than a few percent.  We tested a number of methods for
determining the location of the thermal runaway and found that the most 
sensitive test was the convergence of the code.

The outer boundary condition on the mean UV intensity
is usually set by assuming that there is no incident intensity at the 
boundary (Mihalas 1978).  However, in our situation this is not a valid 
approximation because $\Sigma_2$ (or the Compton depth of the corona)
may be large enough to reflect a significant
amount of radiation back towards the skin.  Instead, we assume
that the UV flux remains constant at $\Sigma < \Sigma_2$ and that $\kappa_{es}
\gg \kappa_{a,\nu}$.  We neglect any contribution from
the hot corona to the Thomson depth at $\Sigma < \Sigma_2$.  We also
neglect (except in the global sense that this is the origin of the X-rays)
any frequency change in the UV photons caused by scattering off gas at
$\Sigma < \Sigma_2$. 
Under these assumptions, there is an analytic relation
between the mean intensity and the flux (Mihalas 1978) which provides our 
boundary condition
\begin{equation}
h_{\nu} J_{\nu} = {1 \over \kappa_{T,\nu}} {\partial (f_{\nu} J_{\nu}) \over 
\partial
\Sigma}
\end{equation}
where
\begin{equation}
\label{eq: surface eddington factor}
h_{\nu} = {1 \over 3} \cdot {1 \over \tau_{2,\nu} + 2/3 
- E_2(\tau_{2,\nu})/3 + E_3(\tau_{2,\nu})/2},
\end{equation}
\begin{equation}
\tau_{2,\nu} = \kappa_{T,\nu} \Sigma_2
\end{equation}
and $E_i$ are exponential integrals.  
In practice, the final value of $\tau_{2,\nu} \ll 1$ and 
eq. \ref{eq: surface eddington factor} reduces to the free streaming value,
$h_{\nu} \simeq 1/\sqrt{3}$.  However, $\tau_{2,\nu} \gtwid 1$ for many
of the intermediate steps.

\section{Results}
\label{sec: results}

\subsection{The Vertical Structure of the X-ray Heated Skin}
\label{subsec: vertical structure}

The inner edge of the X-ray heated skin is 3-4 scale heights from
the midplane of the isothermal disk.
The thickness of the skin increases slightly with
radius but remains $\ltwid 10\%$ of the height of the inner edge (fig.
\ref{fig: structure}a).  The thin 
disk approximation is valid at all radii and the geometry of the
disk can be determined from the isothermal disk solution 
(\S\ref{sec: isothermal disk}).  The density profile in the isothermal
disk is Gaussian so the gas pressure at the inner edge of the skin 
is much smaller (typically by $\sim$ four orders
of magnitude) than the central disk pressure.  This pressure is still
much larger than the pressure in a fully radiation pressure-supported
disk, however.  Throughout most of the skin, 
the radiation force of the UV photons is great enough to balance
gravity and the force of the incident X-ray flux.  The gas pressure is
therefore nearly constant for $\Sigma \ltwid 10
\mbox{gm ${\rm cm^{-2}}$}$ (see fig. 
\ref{fig: structure}b).  
Both the X-ray and UV
fluxes drop to zero at $\Sigma \gtwid 10 \mbox{gm ${\rm cm^{-2}}$}$ and 
a gas pressure gradient forms
to support the disk against gravity.  The gas pressure profile joins smoothly
to the isothermal disk solution at $\Sigma_1$.

The radiative cooling rate is equal to the X-ray heating rate when
the gas is in radiative equilibrium.  Therefore, the gas temperature must
be greater than the UV radiation temperature because the cooling rate is
$\propto T_{g}^4 - T_{r}^4$, where
\begin{equation}
T_{r} = \left( {\pi J \over \sigma} \right)^{1/4}
\end{equation}
and $J$ is the frequency integrated mean intensity of the UV radiation field.  
The actual gas 
temperature profile is determined by three competing effects.  
First, the mean intensity, and $T_{r}$, decrease from the bottom of the skin to
the surface because of the net outward flux.
Second,  the X-ray heating rate increases from zero at the bottom
of the skin to its maximum value at the top.  Therefore,
$T_{r}=T_g$ at $\Sigma_1$ and the difference between 
the two temperatures
increases steadily until $T_g \gg T_{r}$ at
the upper edge of the skin.  Third, the gas temperature gradient changes when
an opacity jump, such as the Lyman edge, becomes transparent (Mihalas 1978, 
p. 206).  The gradient can even change sign in some cases, but not in all cases 
(see fig. \ref{fig: structure}).  The relative importance of these three effects
depends upon the local gas pressure and temperature.  We find
that $T_g$ generally increases from the midplane outward but the 
increase is not always monotonic.

An example of a typical temperature profile 
is presented in fig. \ref{fig: structure}c.  The gas is optically
thick at all frequencies when $\Sigma \gtwid 10 \mbox{gm ${\rm cm^{-2}}$}$ and 
the temperature gradient
\begin{equation}
{\partial T \over \partial \Sigma} \sim 0.
\end{equation}
The gradient is slightly greater than 0 (\ie $T$ increases inward)
at $\Sigma \gtwid 15 \mbox{gm ${\rm cm^{-2}}$}$, and slightly
less than 0 (\ie $T$ increases outward) for $10 \mbox{gm ${\rm cm^{-2}}$}
\ltwid \Sigma \ltwid 15 \mbox{gm ${\rm cm^{-2}}$}$.
The X-ray heating rate increases rapidly at $\Sigma \ltwid 10 
\mbox{gm ${\rm cm^{-2}}$} $ and $T_g$ rises to maintain thermal equilibrium.
In this example, the photosphere for frequencies below the Lyman edge
is at $\Sigma \simeq 0.4 \mbox{gm ${\rm cm^{-2}}$} $, so the temperature
gradient changes sign at this point.  The sign flips back at the photosphere
for frequencies just above the Lyman edge.  The gradient remains
positive at smaller $\Sigma$ and the thermal runaway starts at $\Sigma = 0.012
\mbox{gm ${\rm cm^{-2}}$} $.

The upper boundary of the UV reprocessing region is set by the 
thermal runaway.  The temperature runs away, causing the code to fail to 
converge, when the X-ray ionization parameter
\begin{equation}
\label{eq: ionization parameter}
\Xi = {F_X(\Sigma) \over Pc} \gtwid 1,
\end{equation}
where $F_X(\Sigma)$ is the X-ray flux at $\Sigma$.  When $\Xi$ exceeds
this critical value, bremsstrahlung and recombination cannot cool the gas,
and its temperature rises toward the Compton temperature
(Krolik, McKee \& Tarter 1981; Voit \& Shull 1988; Ferland \& Rees 1988).  
The exact mass column of the thermal runaway ($\Sigma_2$) depends upon
$\dot m$, $m_8$ and $r$, but in general $0.01 \mbox{gm ${\rm cm^{-2}}$}
\ltwid \Sigma_2 \ltwid 1 \mbox{gm ${\rm cm^{-2}}$}$.  $\Sigma_2$ is
never smaller than $\simeq 0.01$ gm~cm$^{-2}$ because the skin is optically
thin to X-rays above that point and the gas pressure doesn't change much
over such a small column density.  On the other hand, $\Sigma_2$
is always less than about one Compton depth because the X-rays
are exponentially attenuated beyond that point.

Typically $\Sigma_2$ rises sharply at larger radii.  A good example of
this effect is seen in the model $m_8=2.7$ and $\dot m = 0.3$ (see fig.
\ref{fig: rsig mdot}).  The reason for this sharp rise is that HeII
recombination provides
much of the cooling at columns $0.1\ltwid \Sigma \ltwid 1$ at
smaller radii, but the lower temperatures at
larger radii mean that little HeIII is formed there.

   Within the framework of the model it is possible to estimate $f_{UV}$,
the fraction of the incident X-ray flux which is reradiated in the UV.
This quantity is determined by how much X-ray flux is either absorbed
or reflected by the material in the thermal runaway layer at column densities
smaller than $\Sigma_2$.  We estimate $f_{UV}$ by the somewhat crude
prescription of assuming that the opacity of the warm layer is unchanged
by the thermal runaway.  A more careful treatment would allow for
the reduction in photoelectric
opacity due to the higher ionization states found there.  Nonetheless,
we believe our global estimate of $f_{UV}$ is not too bad because the
shielding effect is only of major importance when $\Sigma_2$ is at least
$\sim 1$~gm~cm$^{-2}$.  When $\Sigma_2$ is that large, it hardly
matters whether the disk proper is shadowed by photoelectric opacity
or Compton scattering.
 
   The general trend of $f_{UV}$ is to decrease with increasing $\dot m$
because $\Sigma_2$ increases.  The gas pressure in the skin is
\begin{equation}
\label{eq: skin gas pressure}
P_s \sim P_o \left( {1 \mbox{gm ${\rm cm^{-2}}$} \over \Sigma_o} \right)
\propto m_8^{-1}.
\end{equation}
On the other hand, $F_X \propto Q_{dis} \propto \dot m/m_8$ at fixed $r$.
Consequently, $\Xi \propto \dot m$ at the upper boundary of the skin.  
Thus, as $\dot m$ increases, the X-ray flux is able to trigger a thermal 
runaway at larger column densities and over a broader range of radii.
When $\dot m \ltwid 0.003$ the
X-ray flux is never large enough to cause a runaway (see fig. 
\ref{fig: rsig mdot}).

In table \ref{table: uv fraction} we list
\begin{equation}
f_{UV} = {\int rdr F_X(r,\Sigma_2) \over \int rdr F_X(r,0)},
\end{equation}
where $F_X(r,\Sigma)$ is the X-ray flux at a mass column $\Sigma$ in the
ring at radius $r$.  
We find that $f_{UV}$ is 1.0 at $\dot m = 0.003$ and decreases to 0.62
at $\dot m = 0.3$.
The column density of the runaway is insensitive to $m_8$ and $f_{UV}$ varies
only slightly when $m_8$ changes by three orders of magnitude.  We caution
that these numerical values also depend on the specific X-ray
spectral shape we have chosen.

The material above $\Sigma_2$ might influence the emergent
spectrum by radiating soft X-rays and by Comptonizing the UV flux leaving
the disk atmosphere.
We can estimate the gas temperature there as
\begin{equation}
T_{sx} \simeq {{J_{uv} T_r + J_x T_x} \over {J_{uv} + J_x}}
\end{equation}
(Begelman, McKee \& Shields 1988), where $J_{uv,x}$ are the mean intensities of 
the UV and incident X-rays, respectively, and
$T_{r,x}$ are the corresponding Compton temperatures. 
Although we discount the effect of hard X-rays for heating the
disk, they are effective for Compton heating, at least up
to $\simeq 100$~keV, beyond which the Klein-Nishina reduction in the
scattering rate curtails their contribution.  If the spectrum is
$\propto E^{-0.9}$ over many orders of magnitude in photon energy,
$T_x \simeq 3 \times 10^7 \mbox{K}$.  As we have already estimated,
$T_r \ltwid 10^5 \mbox{K}$.  The UV and X-ray intensities are comparable
in the skin, so the additional cooling provided by scattering of the UV photons
will reduce the gas temperature to $T_{sx} \simeq T_x/2 \simeq 1.5 \times 10^7
\mbox{K}$, roughly independent of radius out to the point where the
heating is insufficient to maintain a corona at high temperature.

The emission measure ($EM$) of the soft X-ray corona is dominated by the regions
at larger radius where $\Sigma_2 \sim 1$~gm~cm$^{-2}$.  We estimate 
\begin{equation}
EM \sim 2 \times 10^{65} m_8 r^{1/2} T_5^{1/2} 
\left({\Sigma_2 \over 1 \mbox{gm ${\rm cm^{-2}}$} } \right)^2
\left( {\Delta z \over z_o} \right)^{-1} {\rm cm^{-3}}
\end{equation}
where $\Delta z$ is the thickness of the corona.
We expect $\Delta z \ltwid 0.1 z_o$ and so the
associated bremsstrahlung luminosity
($\sim 10^{43} m_8$~erg~s$^{-1}$) is much smaller than the total disk
luminosity unless $\dot m \ltwid 10^{-3}$. 
Similarly, the importance of Comptonization
to the spectrum depends on the Compton $y$-parameter, $(\tau_T + \tau_T^2)
kT_{sx}/(m_e c^2)$, where $\tau_T$ is the Thomson depth of the layer.
Because $\tau_T$ is at most $\sim 1$ and we expect $kT_{sx}/(m_ec^2)
 \sim 10^{-2}$,
this, too, is unlikely to alter the emergent spectrum.
We elaborate on the effects of Comptonization on the UV spectrum in 
\S\ref{subsubsec: Comptonization}

\subsection{The Reprocessed Ultraviolet Spectrum}
\subsubsection{The Face-On Disk}
\label{subsubsec: face on disk}

  We calculated integrated spectra for face-on disks over a range of
$\dot m$ and $m_8$ chosen to facilitate comparisons with previous
calculations of disk spectra, namely luminosities ranging from 0.003 to
0.3 of the Eddington luminosity and central masses of $2.7 \times 10^7 
M_{\odot}$ to $2.7 \times 10^9 M_{\odot}$.  In each case, we divided the
disk into annuli spaced
at (roughly) equal intervals in $\Delta r / r$ and the local structure and 
spectrum were calculated.  The radii ranged from the
inner radius $r_i = 3.5$
to some outer radius at which the gas
temperature was so low that the contribution
from the outermost radius to the flux near the Lyman edge became negligible.
This occurs at $r \gtwid 20$ where $T_{eff} \ltwid 1.5 \times 10^4 \mbox{K}$
(eq. \ref{eq: teff}).
Finally, the local spectra are added using the method described by LN89, 
which accounts for limb-darkening, Doppler boosting, and relativistic aberration
of the emergent flux.  We neglect the general relativistic bending of light
rays in the gravitational field of the central black hole (\eg Cunningham 1975).
A variety
of face-on spectra are plotted in (figs. \ref{fig: mass spectra},
\ref{fig: mdot spectra}).  Note that we
define $L_\nu (\theta)$ as $2\pi \times$ the luminosity per solid angle
at an angle $\theta$ from the disk axis.

In the simplest treatment of predicted accretion disk spectra,
the dynamics are approximated as Newtonian, and the spectra of individual
rings are assumed to be blackbodies.  These
approximations lead to a temperature profile $\propto r^{-3/4}$
and an integrated spectrum whose shape is roughly $L_\nu
\propto \nu^{1/3}\exp(-h\nu/kT_*)$, where $T_*$ is the temperature
of the innermost ring.  It is clear from these figures that such an
approximation is not a very good description of our predicted spectra.
At low frequencies, there is considerable spectral curvature,
and strong Lyman edge features substantially modify
the shape of the exponential cut-off at high frequencies.

Three different effects alter the spectral shape at frequencies below
the Lyman edge.  First, relativistic effects flatten the temperature profile
in the inner rings (Novikov \& Thorne 1973).  This softens the spectrum
relative to the Newtonian approximation.
Second, the presence of a large emission feature at the Lyman edge implies
that a smaller fraction of the reprocessed flux emerges at
frequencies below the Lyman edge.  This depresses the general level of
the near-UV continuum.
Third, at frequencies below the Lyman edge, the mass column of the UV
photosphere increases with increasing frequency.  The flux at lower
frequencies then comes from hotter gas because
the gas temperature typically increases as $\Sigma$ decreases.  This softens
the continuum spectrum relative to the black body approximation.

The low-frequency spectrum becomes harder with increasing hard X-ray flux,
\ie $\dot m/m_8$. 
The disk temperature is roughly proportional to the effective temperature of
the X-ray flux, so the peak of the disk spectrum shifts from $10^{15}$ Hz
in the coldest disks to $\gtwid 10^{15.5}$ Hz in the hottest.
To quantify this effect, we define $\alpha_{15} = d\log L_\nu/d\log\nu$
at $\log\nu = 15.0$.  The value of this quantity for each model
we computed is displayed in table \ref{table: uv fraction}.
The smallest $\alpha_{15}$ we found was -1.6 ($\dot m m_8^{-1} \sim
10^{-3}$); the largest was 0.8 ($\dot m m_8^{-1} \sim 10^{-1}$).

  As the previous paragraphs explain, some, but not all,
these effects are due to the external irradiation.  In particular, the
temperature gradient effect acts the opposite way when the dissipation
is inside the disk rather than outside. 
An improved calculation of the spectrum of a conventional
disk with internal dissipation methods yields a predicted
$\alpha_{15} \simeq 1$ (Sincell \& Krolik 1996).

   We found a strong Lyman edge emission emission feature over the
entire parameter range we studied (see figs. 
\ref{fig: localspec2}, \ref{fig: mass spectra} and \ref{fig: mdot spectra}).
This, too, is due to the temperature increasing upward (\ref{fig: structure}d,
\ref{fig: structure}c).  The larger opacity
above the edge than below causes the Lyman continuum photosphere to lie
higher in the atmosphere than the photosphere at frequencies just below
the edge.  The higher temperature at the Lyman continuum photosphere naturally
leads to a greater emergent flux, i.e. an emission feature (fig. \ref{fig:
localspec2}).  Additional emission features appear at the HeI and HeII
edges in some of the integrated spectra.  The smearing of the edges is caused
by Doppler boosting of radiation emitted at small $r$.

The amplitude of the Lyman edge feature increases with the central mass
(fig. \ref{fig: mass spectra}) because $P_s \propto m_8^{-1}$ (eq.
\ref{eq: skin gas pressure}).
Low pressure gas cools less efficiently because both the bremsstrahlung and
photoionization opacities are proportional to $P_s$.  Inefficient cooling leads
to larger temperature inversions in the skin. The ratio $\Sigma_{Ly+}/
\Sigma_{Ly-}$ remains roughly constant because the ratio of the opacities
is independent of $P_s$.  Therefore, $T(\Sigma_{Ly+}) /
T(\Sigma_{Ly-})$ increases with $m_8$ and the amplitude of the Lyman edge
feature increases.
The amplitude of the edge feature is nearly independent of $\dot m$ (fig.
\ref{fig: mdot spectra}).

\subsubsection{Angular Distribution of the Emergent Flux}
\label{subsubsec: angular distribution}

We incorporated an iterative calculation of the variable Eddington factors
(Auer \& Mihalas 1970) 
into the $\dot m=0.03$, $m_8=2.7$ model.
The solution for the Eddington factors ($f_{\nu}$) 
and the mean intensity 
is equivalent to the exact solution for the angular distribution of the
spectral intensity.  The angular distribution of the emergent flux can be
determined by a formal solution of the transfer equation with the 
known source function.
The surface Eddington factors (eq. \ref{eq: surface eddington factor}) cannot
be computed iteratively because we do not have a solution at 
$\Sigma < \Sigma_2$.

The accuracy of the analytic approximation for the Eddington factors
(eq. \ref{eq: analytic fedd}) was tested by comparison with the exact solution.
The radiation field is isotropic when $\tau \gg 1$ and $f_{\nu} = 0.33$, 
in good agreement with the analytic approximation.
However, when $\tau \ll 1$ we find that $f_{\nu} \ltwid 0.5$.  The approximate
$f_{\nu}$ assumes that the radiation is collimated perpendicular to the
accretion disk so that $f_{\nu} = 1$.
Fortunately, the errors introduced by our assumed $f_{\nu}$ are small.
The approximate solution underestimates the
true gas temperature by $\ltwid 5\%$ at $\Sigma \ltwid 1 
\mbox{gm ${\rm cm^{-2}}$} $ and overestimates the temperature by $\ltwid 10\%$
at larger 
$\Sigma$.  The gas pressure is underestimated by $\ltwid 10\%$ at all $\Sigma$.
The differences in the UV spectrum depend on frequency and never exceed $10\%$.
The approximate solution overestimates the mean intensity below the Lyman edge
and above the HeII edge and underestimates $J_{\nu}$ between these frequencies.
As a consequence, the amplitude of the Lyman edge feature is larger in the exact
solution.

The observed flux from an accretion disk around a massive black 
hole depends upon the disk inclination angle.
First, as discussed in the previous paragraph, the emergent flux has an
intrinsic angular distribution in the local rest frame (see below).
Second, there are a number of relativistic effects acting upon photons
emitted from material orbiting near the black hole.  Relativistic
orbital motion boosts their frequencies and collimates their directions,
while the strong gravitational field bends their trajectories and imposes
a gravitational redshift on their frequencies as viewed at large distance
(Cunningham 1975).
The boosting and beaming due to orbital motion tend to be stronger than 
the gravitational effects, so the net result is to strengthen the
radiation emitted close to the equatorial plane from
the innermost rings of the disk (Cunningham 1975, Laor, Netzer \& Piran 1990).
These effects are stronger in rotating black holes than in non-rotating
black holes because the disk extends closer to the event horizon in
the former case (Laor, \etal 1990).  
For this reason, the impact of general relativistic effects
is small in our (non-rotating black hole) models. 

We calculated the integrated spectrum of the $\dot m = 0.03$, $m_8=2.7$ disk
for three disk inclination angles.
For each inclination, the disk is divided into several annuli (\S
\ref{subsubsec: face on disk}) and each annulus subdivided into  several 
sectors.  The radiative transfer equation is solved for the spectral intensity 
at the appropriate rest-frame inclination
angle using the exact source function and the results are summed over all
annuli and sectors.  As before, Doppler boosting, relativistic aberration
and the gravitational
redshift are included, but we neglect the general relativistic 
bending of the light rays
by the central black hole.

At low frequencies, the emergent intensity is limb-darkened (fig.
\ref{fig: limb disk}).  Due to
the low opacity below the Lyman edge, the photosphere for these
frequencies tends to lie relatively deep inside the atmosphere,
where the temperature gradient is small or positive (i.e. temperature
decreases upward).  Consequently, the angular dependence is determined
largely by the usual limb-darkening influence of scattering.
However, at frequencies well above the Lyman edge, the intensity is more
nearly isotropic.  Several effects combine to explain this change.
Because of the greater opacity, the photosphere in this range lies
higher in the atmosphere, where the temperature tends to increase outward.
Therefore, photons traveling well away from the axis can only depart
from relatively high altitude, where the temperature is comparatively high.
In addition, because these frequencies are radiated almost exclusively by
the innermost rings of the disk, relativistic boosting and beaming
are especially important.  The decreasing degree of limb-darkening with
increasing frequency causes the
feature at the Lyman edge to strengthen with disk inclination.

We expect the relative limb-brightening of the ionizing continuum to
increase with central mass and, if $L/L_E$
is approximately constant, disk luminosity.  This is
because temperature gradients increase with $m_8$ (\S
\ref{subsubsec: face on disk}).

Doppler broadening of the Lyman edge increases with disk inclination.  The
typical line of sight to an edge-on disk is nearly parallel to the disk 
rotation velocity so the Doppler boost is maximized.   The Lyman edge is 
formed at $r\sim 10-20$, where $\beta_{\phi} \sim 0.2$, so the edge is 
spread over a range ${\Delta \nu \over \nu} \ltwid 0.2$ (fig.
\ref{fig: limb disk}).  The peak of the emergent flux shifts upward by the 
same amount.   An edge-on disk will not appear to have an emission feature
at the Lyman edge, but the continuum slope becomes harder going from
frequencies below the Lyman edge to higher frequencies.
This result is consistent with the conclusions of Laor (1992).

\subsubsection{Comptonization of the UV Radiation} 
\label{subsubsec: Comptonization}

The reprocessed UV radiation is inverse Compton scattered by the relativistic 
electrons in the coronae.
We calculate the Comptonized UV spectrum for scattering by a relativistic 
Maxwell-Boltzmann distribution of electrons
\begin{equation}
N(\gamma) = {1 \over \Theta K_2(1/\Theta)} \gamma (\gamma^2-1))^{1/2}
\exp(-\gamma/\Theta),
\end{equation}
where $\Theta = kT_c/m_e c^2$, the coronal temperature is $T_c$
and $K_2$ is the second order modified Bessel function.
This distribution is normalized so that
\begin{equation}
\int d\gamma N(\gamma) = 1
\end{equation}
and we assume a Thomson depth, $\tau_c$, for the corona.
We further assume that each photon has a probability $\tau_c < 1$ of being 
scattered in the corona and that each photon is scattered at most one time.
The spectrum of the scattered radiation is calculated using the formulae for
isotropic electron and photon distributions (Rybicki \& Lightman 1979).
The small corrections due to anisotropy of the incident UV radiation 
field were neglected 
(Haardt 1993).
Higher order scatterings determine the shape of the high energy spectrum
(Pozdnyakov, Sobol \& Sunyaev 1976) so we plot the results over a limited
range of frequencies near the Lyman edge (figs. \ref{fig: Compton t} and 
\ref{fig: Compton tau}).

The spectrum of the Comptonized UV flux is determined primarily 
by the Thomson depth of the corona, $\tau_c$.
Inverse Compton scattering in an optically thin corona has a negligible
effect on the spectrum of the UV continuum because few photons are
scattered (fig. \ref{fig: Compton t}).  
Comptonization in an optically thick corona ($\tau_c \gtwid 0.5$) smooths 
the spectrum at all 
frequencies (fig.
\ref{fig: Compton tau}), preferentially scattering photons from 
low to high energy.  
As a consequence, the
spectrum of the non-ionizing flux becomes steeper and the ionizing
continuum hardens to an approximate power-law.  

Because the width of the Lyman edge is dominated by relativistic boosting,
inverse Compton scattering in an optically thick corona may reduce its
amplitude but does not significantly broaden it (fig. \ref{fig: Compton tau}).
The amplitude of the emission feature reaches a minimum when $0.1 \ltwid
\Theta \ltwid 1$.  The actual value of $\Theta_{min}$ depends upon the 
parameters of the corona and the input UV spectrum but the range of values is 
easily understood.  Each input photon
has a probability $\sim \tau_c$ of being scattered in the corona and the
scattered photons are redistributed over a frequency range ${\Delta \nu \over
\nu} \sim \Theta$.  Thus, very little redistribution occurs when $\Theta \ll
1$ and the scattered photons emerge at much higher energies when $\Theta \gg 1$.

The total optical depth of the corona is approximately
the sum of the contributions from the soft X-ray corona
(discussed in \S\ref{subsec: vertical structure}) and the hard X-ray corona, 
where the gravitational potential energy is released.
In \S\ref{subsec: vertical structure} we demonstrated that a thermal runaway
occurs for 
$\Sigma_2 \sim 0.1 \mbox{gm ${\rm cm^{-2}}$} $ for $r \ltwid 50$
and 
$\Sigma_2 \sim 1 \mbox{gm ${\rm cm^{-2}}$} $ at larger radii.
This range of column densities corresponds to $\tau_{sx} \sim 0.04 - 0.4$.  The
Lyman edge is formed at $r \sim  10-20$, where $\tau_{sx} \ltwid 0.04$.
The optical depth ($\tau_{hx}$) and geometry of the hard X-ray corona 
are controversial, but may be estimated through the use of thermal
Comptonization models.  HM93 find $\tau_{hx} \ltwid 0.5$ assuming a uniform
slab; Zdziarski \etal (1995) found $\tau_{hx} \simeq 0.08$ for the same
geometry; Pietrini \& Krolik (1995) confirmed the Zdziarski \etal (1995) model,
but also suggested that either clumping of the corona or a physical separation
between it and the reprocessing surface would permit a match to the observed
X-ray slope with larger $\tau_{hx}$.  Stern, \etal (1995) prefer
a model in which the corona is clumped, but close to the accretion disk,
and has $\tau_{hx} \ltwid 0.1$. 

   If the corona is highly clumped, it can have little effect on the
bulk of the UV photons.  These modeling efforts indicate that a smooth
corona must be rather optically thin, so it seems unlikely that the
mean total Compton depth overlying the rings producing the Lyman
edge can be any greater than a few tenths in most cases.

\section{Gravitational Instability of the Cold Disk}
\label{sec: gravitational instability}
Accretion disks are self-gravitating and 
unstable when (Toomre 1964, Goldreich \& Lynden-Bell 1965, 
Sakimoto \& Coroniti 1981)
\begin{equation}
Q = {z_o \Omega_k^2 \over G \Sigma_o} < 1,
\end{equation}
or
\begin{equation}
\label{eq: self gravity criterion}
\rho_o \gtwid {M \over 4\pi R^3} = 0.617 m_8^{-2} r^{-3}
\mbox{gm ${\rm cm^{-3}}$},
\end{equation}
where $\Omega_k$ is the Keplerian angular frequency.  The range of 
length scales for the growing perturbations is
$Q \ltwid l/z_o \ltwid 1/Q$ (Toomre 1964).
Comparing eqs. \ref{eq: isothermal gas density} and 
\ref{eq: self gravity criterion}, we see that an isothermal accretion disk
with no internal dissipation is gravitationally stable if
\begin{equation}
\label{eq: isothermal disk instability}
\dot m m_8 T_5^{-3/2} \alpha_{SS}^{-1} \ltwid 0.9.
\end{equation}

The temperature of
the isothermal disk is reasonably well approximated by the effective temperature
of the irradiating flux (eq. \ref{eq: teff}) because the Thomson depth of the
skin is near unity.  In contrast, the central temperature of the standard 
optically thick disk
can be more than an order of magnitude larger than $T_{eff}$.  
Inserting $T_{eff}$ into
the stability criterion (eq. \ref{eq: isothermal disk instability}), 
we find that
the isothermal disk is unstable at radii
\begin{equation}
r \gtwid 2 \dot m^{-5/9} m_8^{-11/9}.
\end{equation}
The disk is self-gravitating at all radii ($r > 3$) when
\begin{equation}
\label{eq: grav luminosity condition}
L \gtwid 3 \times 10^{45} m_8^{-11/5} 
\mbox{ ergs ${\rm s^{-1}}$}.
\end{equation}
Of the disks we have considered, only the $\dot m = 0.03$ and $m_8 = 0.27$ model
satisfies this criterion over the entire range of radii.  The case $\dot m = 
0.003$ and $m_8 = 2.7$ is stable for $r \ltwid 10$ and the 
remainder are unstable
at all radii.

Although a strictly isothermal disk is unstable, a very small amount of 
internal dissipation will stabilize an irradiated disk.
The gas pressure dominated solution for the disk structure described in SZ95 
applies when there is enough internal dissipation (and optical depth)
to significantly elevate the central temperature above $T_{eff}$ (SZ95 and
appendix \ref{app: cold disk structure}).   This occurs when 
\begin{equation}
\label{eq: tdiff criterion}
(1-f) \gg {8 \over 3\tau_d} (1-f/2),
\end{equation}
where $\tau_d$ is the electron scattering optical depth of the disk and
$f$ is the fraction of the gravitational potential energy which is
dissipated in the corona (appendix \ref{app: cold disk structure} and SZ95).
As a result of the warmer internal temperature, the scale height increases
and the density falls.  We find that the central density drops below the
critical density (eqs. \ref{eq: isothermal gas density} and \ref{eq: self
gravity criterion}) and the disk is stabilized when
\begin{equation}
\label{eq: stability criterion}
(1-f) \gtwid 5 \times 10^{-16} \alpha_{SS}^{-7/3} m_8^{13/3} \dot m^{4/3}
r^{9/2}.
\end{equation}
All of our models are stable at $r\ltwid 50$ with small values of
$1-f$.  The largest value of $1 - f$ required for stability occurs
in the $\dot m = 0.03$ and $m_8 = 27$ case at $r = 50$.  Substantially smaller
values of $1 - f$ are needed to stabilize the 
disk at $r \ltwid 10$, where most of the Lyman edge radiation is produced, and
for smaller values of $\dot m$ and $m_8$.
In general, $4\pi R^3 \rho_o /M$ increases, 
and the value of $1 - f$ required to stabilize the disk decreases, with
increasing luminosity and $m_8$ (eq. \ref{eq: self gravity criterion}). 

    From the scaling behavior we have already discovered, we may make some
educated guesses about how the disk spectra will change if $1-f$ changes
from zero to the small values required for stability.
Because the density throughout the disk will fall, while X-ray heating will
keep the gas temperature in the heated layer near $T_{eff}$, we expect
the gas pressure to decrease with increasing $1-f$.  Therefore, the ionization
parameter in the heated skin will increase, and more of the
incident X-ray flux will emerge as soft X-rays rather than in the ultraviolet.
(\S\ref{subsec: vertical structure}).  
In addition, we expect the amplitude
of the Lyman edge feature will increase as $1 - f$ increases because of
the lower gas pressure in the atmosphere (\S\ref{subsubsec: face on disk}).

The disk is supported by radiation pressure at $r\sim 6$ when 
\begin{equation}
f \simeq 1 - 0.05 \dot m^{-8/9} \left(\alpha_{SS} m_8 \right)^{-1/9}
\end{equation}
and over a larger range of radii for smaller $f$ (SZ95).
Radiation pressure support will inflate the disk, suppressing the
gravitational instability,  but may cause the disk to become
thermally unstable (\eg Piran 1978).  
However, for our range of model parameters, the value of $f$ needed to
stabilize the disk is much smaller than this value
in all but the highest luminosity disk ($\dot m = 0.3$ and
$m_8=2.7$).

\section{Conclusions}
\label{sec: conclusions}

We have calculated the vertical structure and UV spectrum of an
accretion disk irradiated by hard X-rays from a corona, as predicted
by many models for the X-ray emission in AGN (\eg Pounds \etal 1990;
HM93). Consistent with these X-ray production models, we assume
there is no viscous
dissipation in the accretion disk.  The gravitational
potential energy of the accreting gas is dissipated in a corona and
radiated as hard X-rays with a power law spectrum.  Photoionization
of oxygen and iron by the coronal X-rays heat a geometrically thin
skin at the surface of the accretion disk.  Because there is no dissipation
inside the disk proper, its temperature is constant as a function of
altitude and the gas density and optical depth are much larger than in
a conventional dissipative disk (SZ95).  

To find the structure of the X-ray heated skin, we have solved the 
differential equations of radiative transfer, hydrostatic equilibrium
and surface mass density subject to the constraints of radiative equilibrium
and charge conservation.  The inner boundary conditions for this numerical
solution were chosen to match an analytical solution for the 
underlying isothermal accretion disk.
We tested our method by computing the structure and spectrum of a conventional
accretion disk and comparing to the previous state-of-the-art calculation
(LN89, Sincell \& Krolik 1996).

The inner edge of the X-ray heated skin is 3-4 scale heights from the 
midplane of the isothermal disk.  In the skin, but not the body of the disk, the
force of the UV flux supports the gas against gravity and the
momentum of the absorbed X-rays.  The gas pressure in the skin is orders
of magnitude smaller than the pressure at the midplane of the disk and is
nearly constant throughout the skin.  The gas temperature typically 
increases from the
inner edge of the skin to the surface, although the temperature 
gradient occasionally changes sign.  The changes in the temperature gradient
correspond to changes in the ionization state and opacity of the gas. 

 The gas above the column
density $\Sigma_2$ is unable to cool because cooling by thermal radiation 
processes is no longer able to balance heating by X-ray
photoionization.  This transition takes place at a characteristic
value of the ionization parameter $\Xi \simeq 1$.  Compton
cooling was not included in this model so we could not find a solution for
$\Sigma < \Sigma_2$.  However, 
we estimate the gas temperature in the soft X-ray corona to be $T_{sx} 
\simeq 1.5 \times 10^7 \mbox{K}$.
We find that $\Sigma_2$ increases with disk radius, because the cooler gas 
at large radii
radiates less efficiently, and the average value of $\Sigma_2$ for a given
disk increases with
the accretion rate, because
$\Xi \propto \dot m$ at the surface of the heated skin.

For our specific X-ray spectrum and set of approximations,
the fraction of the incident hard X-ray
flux which is reprocessed into
UV photons ranges from $f_{UV} = 1.0$ at $\dot m = 0.003$ to 0.62 
when $\dot m = 0.3$.  The exact value of $f_{UV}$ depends on the shape of the 
X-ray spectrum, and our method (see \S 4.1) can be expected to slightly
underestimate $f_{UV}$.
Nevertheless, the decrease in $f_{UV}$ with accretion rate is a direct 
consequence of our assumption that all the gravitational potential energy is
dissipated in the corona (see eq. \ref{eq: skin gas pressure})
and should be  a general property of irradiated disks.

The reprocessed UV continuum depends significantly on the specific parameters
of the model (accretion rate, central mass), but certain features are
generic: it has strong spectral curvature, in the sense that the spectrum
softens with increasing frequency; and there is a significant Lyman edge
emission feature. The low frequency continuum (parameterized
by the spectral index at $10^{15} \mbox{Hz}$) hardens with increasing
X-ray flux (\ie $\dot m/m_8$) and ranges from -1.6 when $\dot m m_8^{-1}
\sim 10^{-3}$ to 0.8 when $\dot m m_8^{-1} \sim 10^{-1}$.  The continuum
slope in this frequency range is only weakly dependent on viewing angle. 

   The Lyman edge is in emission because the photosphere is higher
in the atmosphere at frequencies above the edge, and the temperature
increases upward.
Additional emission features at the HeI and HeII edges may
also occur in the spectra of the hotter disks.
The amplitude of the Lyman edge feature increases with central mass and is
insensitive to the accretion rate.  This is because both the photoionization and
bremsstrahlung opacities are proportional to $P_s \propto m_8^{-1}$, so the
gas cools less efficiently as $m_8$ increases.  This results in larger 
temperature inversions which increase the strength of the emission feature.
Doppler boosting spreads the Lyman edge over a range ${\Delta \nu / \nu} \sim
\beta_{\phi} \sim 0.2$.
The amplitude of the Lyman edge increases with disk inclination because
the non-ionizing continuum is limb-darkened and the ionizing flux is
(relatively) limb-brightened.

Irradiated disks are self-gravitating, and unstable, unless there is a small
amount of internal dissipation.
We find that the strictly isothermal disk is 
gravitationally unstable 
at $r \gtwid 2 \dot m^{-5/9} m_8^{-11/9}$, so that strictly isothermal
disks with luminosities $L \gtwid 3 \times 10^{45} m_8^{-11/5}$~ergs~s$^{-1}$
are unstable at all radii.
If, as is likely, there is a small amount of internal dissipation,
and this stabilizes the disk, we expect 
$f_{UV}$ to decrease, and the amplitude of the Lyman edge feature to 
increase relative to what our models show for the limiting case of $f=1$.
These effects are likely to become stronger for larger luminosity and central
mass.

   Some, but not all, of these predictions are consistent with observations.
The average optical-UV power-law index for quasars is $\sim -0.5$ (Laor 1990)
but a single power law is a poor description of the continuum spectrum
(Francis, \etal 1991).
The continuum of the composite quasar spectrum, formed by combining all the
quasars in the Large Bright Quasar Survey, softens with increasing frequency
(Francis, \etal 1991).
Both the mean optical-UV slope and the spectral curvature are consistent
with our predicted UV spectrum of an irradiated
disk if $\dot m m_8^{-1} \sim 10^{-2}$. 

  On the other hand, our finding that a Lyman edge emission feature is a
general property of
X-ray-irradiated disks is difficult to reconcile with recent observations
of AGN.  Lyman edge features are rarely seen, and then usually as partial
absorption edges (Antonucci, Kinney \& Ford
1989, Koratkar, Kinney \& Bohlin 1992).   Emission features at the Lyman
edge have been detected only in polarized light, and in these
two quasars the total flux spectrum shows the Lyman edge in 
absorption (Koratkar, \etal 1995).  The polarization of this emission
is quite high and detailed polarization dependent calculations (\eg Blaes
\& Agol 1996) will be
required to determine if these features could be produced by an irradiated 
disk.  

Internal dissipation might, in some cases, reduce the amplitude of the 
Lyman edge feature.  Conventional disks (Sincell \& Krolik 1996)
sometimes produce emission features at the Lyman edge, and sometimes
absorption, depending primarily on $\dot m$.  However, both the fraction
of the total gravitational energy dissipated in the disk and $\dot m$
must be very finely tuned to eliminate Lyman edge features from all AGN.

Inverse Compton scattering in the disk coronae can smooth out the Lyman
edge feature
only if the hot corona producing the hard X-rays covers most of the disk
and has an optical depth $\tau_c \gtwid 0.5$.  Because most recent
treatments of X-ray production suggest that either the optical depth is
rather smaller than this, or else the hot corona is highly clumped, 
it is unlikely that Compton scattering smooths the Lyman edge.

The only escape we see from this contradiction 
makes use of the
fact that the evidence for X-ray irradiation comes almost entirely from
low luminosity AGN (\ie Seyfert galaxies), while spectra of the Lyman
edge region exist only for high luminosity AGN (\ie quasars).  The reason
for this separation is that only the Seyfert galaxies, which are relatively
nearby, are bright enough for detailed X-ray spectroscopy, while in order
for us to see the Lyman edge, we must look at objects with redshifts great
enough to bring it below the Lyman cut-off imposed by our Galaxy's
interstellar medium.  Perhaps, then, Seyfert galaxies have Lyman edges
in emission (if we could only see them), while quasar disks are not so
thoroughly irradiated by X-rays.  X-ray spectroscopy of quasars by future,
more sensitive instruments, will be able to test this idea.

   Thus, our calculation of the UV spectrum expected from irradiated disks
raises serious questions for accretion disk
models.  While the X-ray production and X-ray reflection models which
motivated this study have had substantial success in describing AGN
X-ray spectra, we now see that they also predict strong Lyman edge emission
features, a prediction that is contradicted by the simplest interpretation
of the observations.  

\acknowledgements
We would like to thank Ari Laor for helpful conversations and for providing 
many disk spectra.  We also thank Omer Blaes and Eric Agol for 
ongoing conversations.  
MWS also thanks the Observatoire de Meudon for hospitality
during part of this work.  MWS received support for this work from NASA
grants NAGW-3129, 1583 and NAG 5-2925, and NSF grant AST 93-15133.  JHK
was partially supported by NASA Grant NAGW-3156.

\begin{appendix}

\section{Radial Structure of the Cold Disk for $f \simeq 1$}
\label{app: cold disk structure}
The radial structure equations for the cold gas pressure dominated disk
are (SZ95)
\begin{equation}
P = {GM \rho \over 2 R^3} z_o^2 = {\rho c^2 \over 4} r^{-1} 
\left( {z_o \over R} \right)^{2},
\end{equation}
\begin{equation}
P = {\dot M \omega \over 4\pi\alpha_{SS}} {\sqrt{\pi} \over z_o}
= \sqrt{ {\pi \over 8} } {m_p c^2 \over \sigma_T R_s} r^{-5/2}
  {\dot m \over \alpha_{SS} \eta} \left( {z_o \over R} \right)^{-1}
\end{equation}
and the perfect gas equation of state.  The gas temperature of the 
disk is determined by the X-ray heating rate (eq. \ref{eq: dissipation rate})
\begin{equation}
T \simeq \left( {(1-f/2) Q_{dis} \over \sigma} \right)^{1/4}
\end{equation}
when $f \simeq 1$.  SZ95 assume that the gas temperature is proportional to 
the effective temperature of the flux dissipated {\it internally} and neglect 
the X-ray heating.  
Consequently, the SZ95 solution for the radial structure is singular
($\rho \rightarrow \infty, T = 0$) when $f=1$.

The solution for the radial structure is straightforward.  We find that the
scale height is
\begin{equation}
z_o^2 = {2 k T_{eff} R^3 \over GMm_p} (1-f/2)^{1/4},
\end{equation}
where
\begin{equation}
\label{eq: teff}
T_{eff} = \left( {Q_{dis} \over \sigma} \right)^{1/4}
\end{equation}
The midplane gas density is
\begin{equation}
\rho = 0.53 \dot m m_8^{-1} \alpha^{-1} r^{-3} T_{eff}^{-3/2} (1-f/2)^{-3/8}
\end{equation}
where we have inserted the scalings in eq. \ref{eq: disk parameter scalings}.
These equations reduce to the isothermal disk solution 
(eqs. \ref{eq: disk scale height}
and \ref{eq: isothermal gas density}) for a disk temperature of $T_g = T_{eff}$
when $f=1$, as expected.

The introduction of a new definition of the disk temperature removes the 
singularity of the SZ95 solution at $f=1$.  The cold  disk temperature will
never fall to zero, as predicted by SZ95, because of the X-ray heating by
the corona.  As a result, the scale height is non-zero and the density
remains finite for $f=1$.  In addition, we find that the scale height and 
midplane density of a gas pressure dominated disk are insensitive to the actual
value of $f$, so long as it is close enough to unity that the central
temperature is determined primarily by irradiation ({\it cf.} \S 5).

\end{appendix}

\begin{figure}
\centerline{\plotone{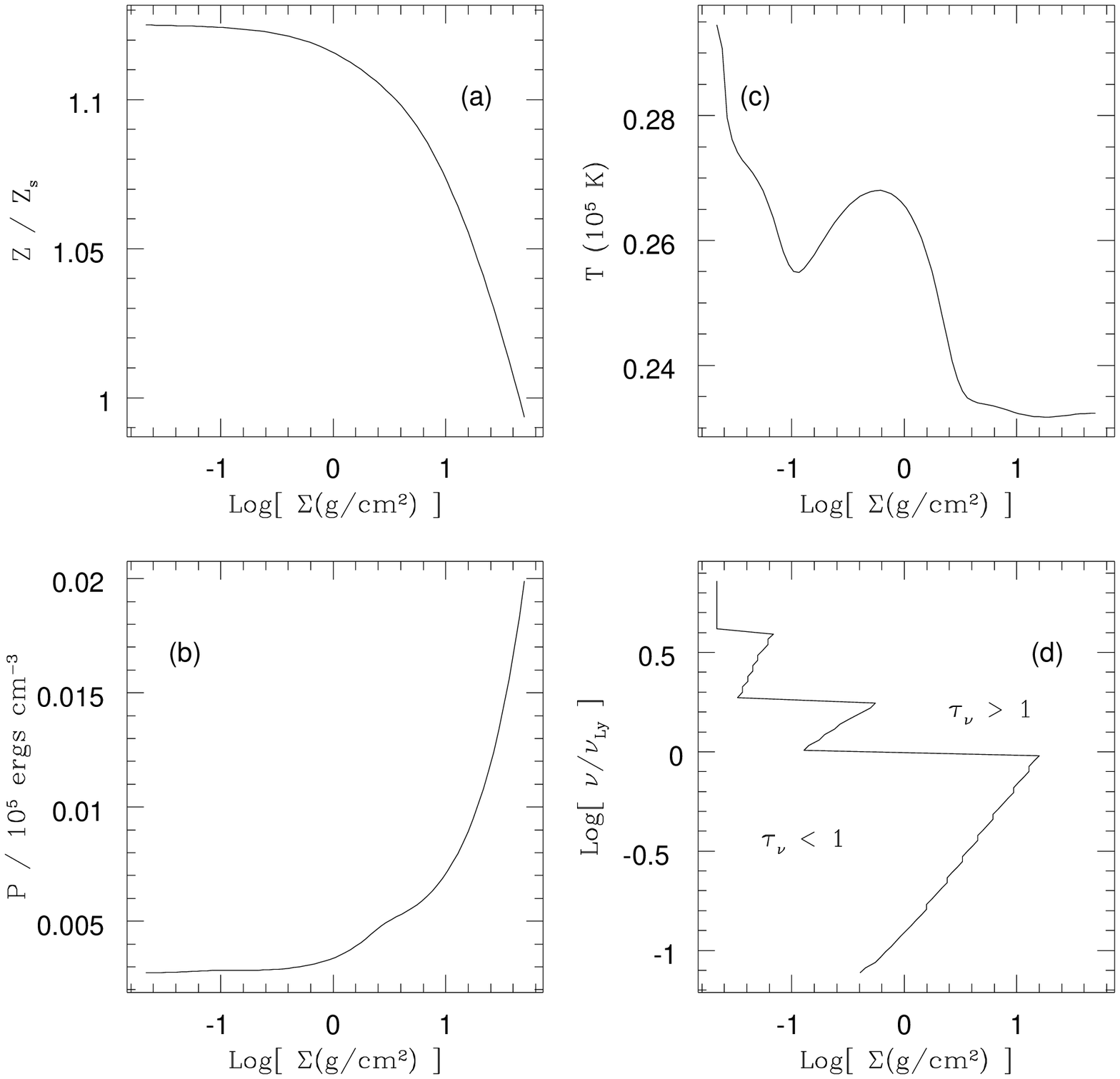}}
\caption{\label{fig: xray heating rate} }
\end{figure}

\vfil\eject
\centerline{\bf Figure and Table Captions}
\bigskip
Figure \ref{fig: xray heating rate}.  The X-ray heating rate and integrated
UV flux as a function
of column density.  Units are arbitrary.

\begin{figure}
\centerline{\plotone{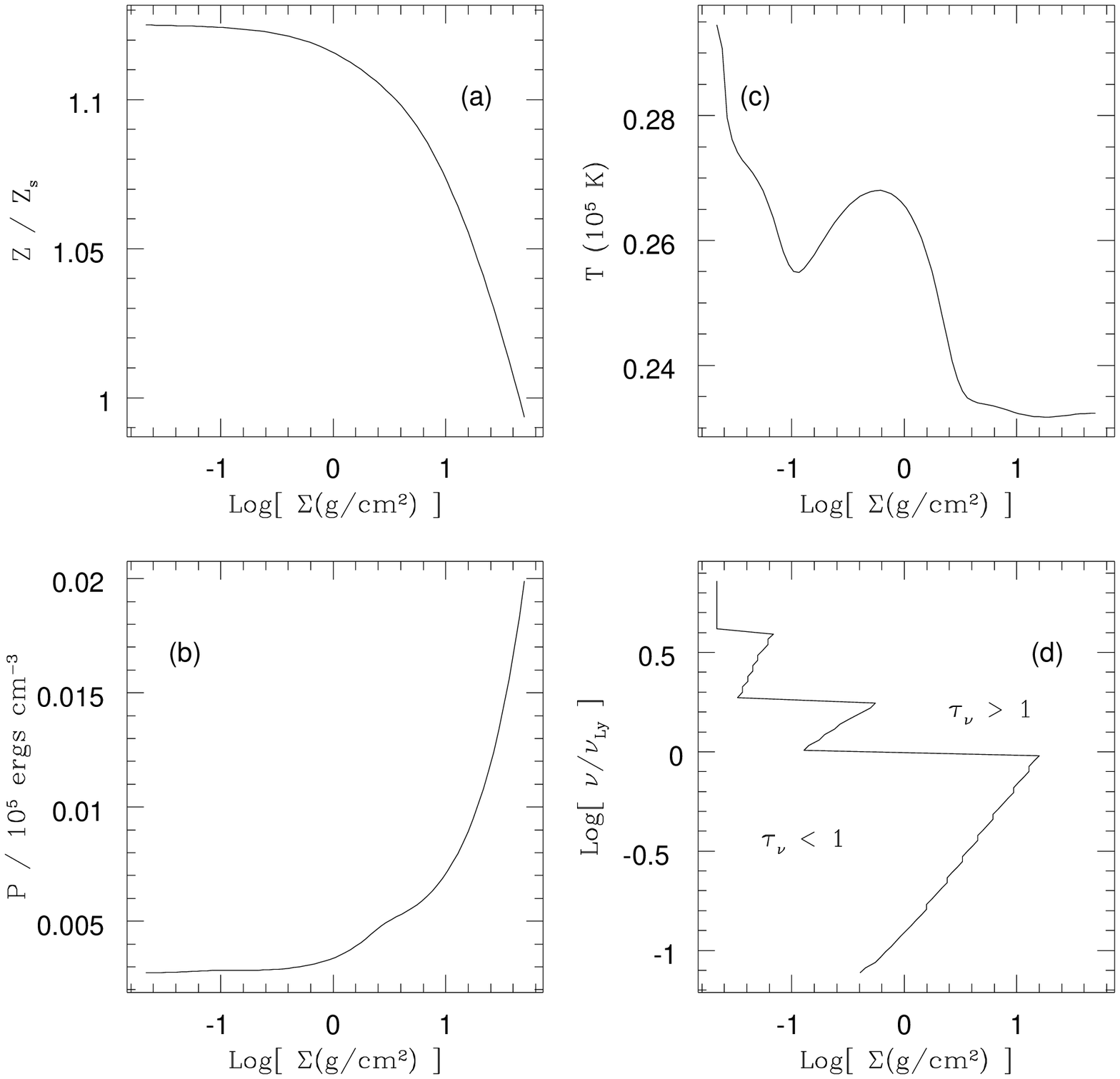}}
\caption{\label{fig: structure} }
\end{figure}

\bigskip
Figure \ref{fig: structure}. The structure of the X-ray heated skin at $r=14.0$
for $\dot m = 0.03$ and $m_8=2.7$.  The panels show the vertical coordinate
scaled to the value at the inner edge of the skin ($Z_s$) (a), the
gas pressure (b), the gas temperature (c) and 
the location of the photosphere (d). 

\begin{figure}
\centerline{\plotone{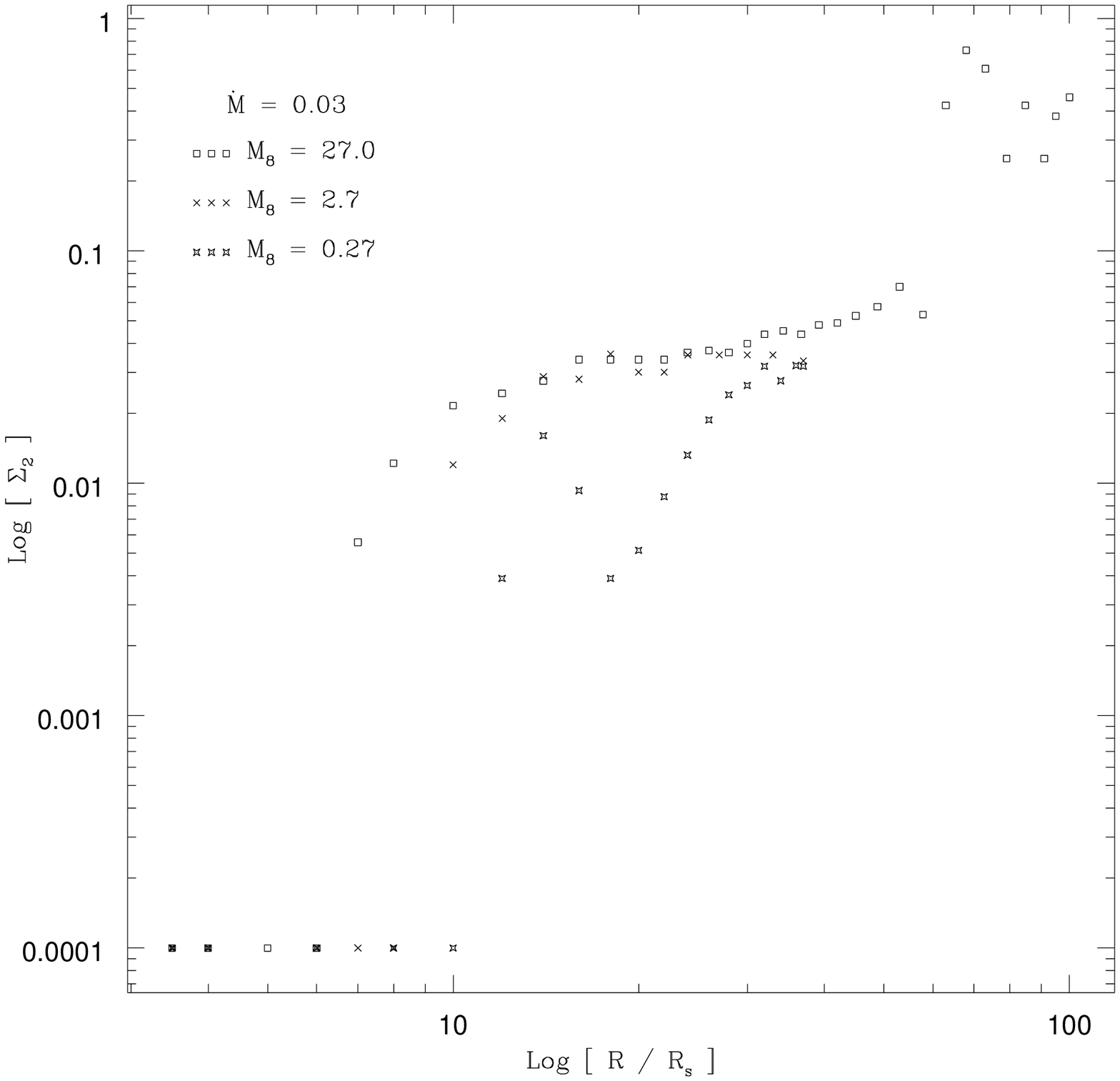}}
\caption{\label{fig: rsig mass} }
\end{figure}

\bigskip
Figure \ref{fig: rsig mass}.  The mass column of the thermal runaway 
($\Sigma_2$) for fixed $\dot m$ and variable $m_8$.  
The outer boundary
of the integration is $\Sigma = 10^{-4} \mbox{g ${cm^{-2}}$}$. 

\begin{figure}
\centerline{\plotone{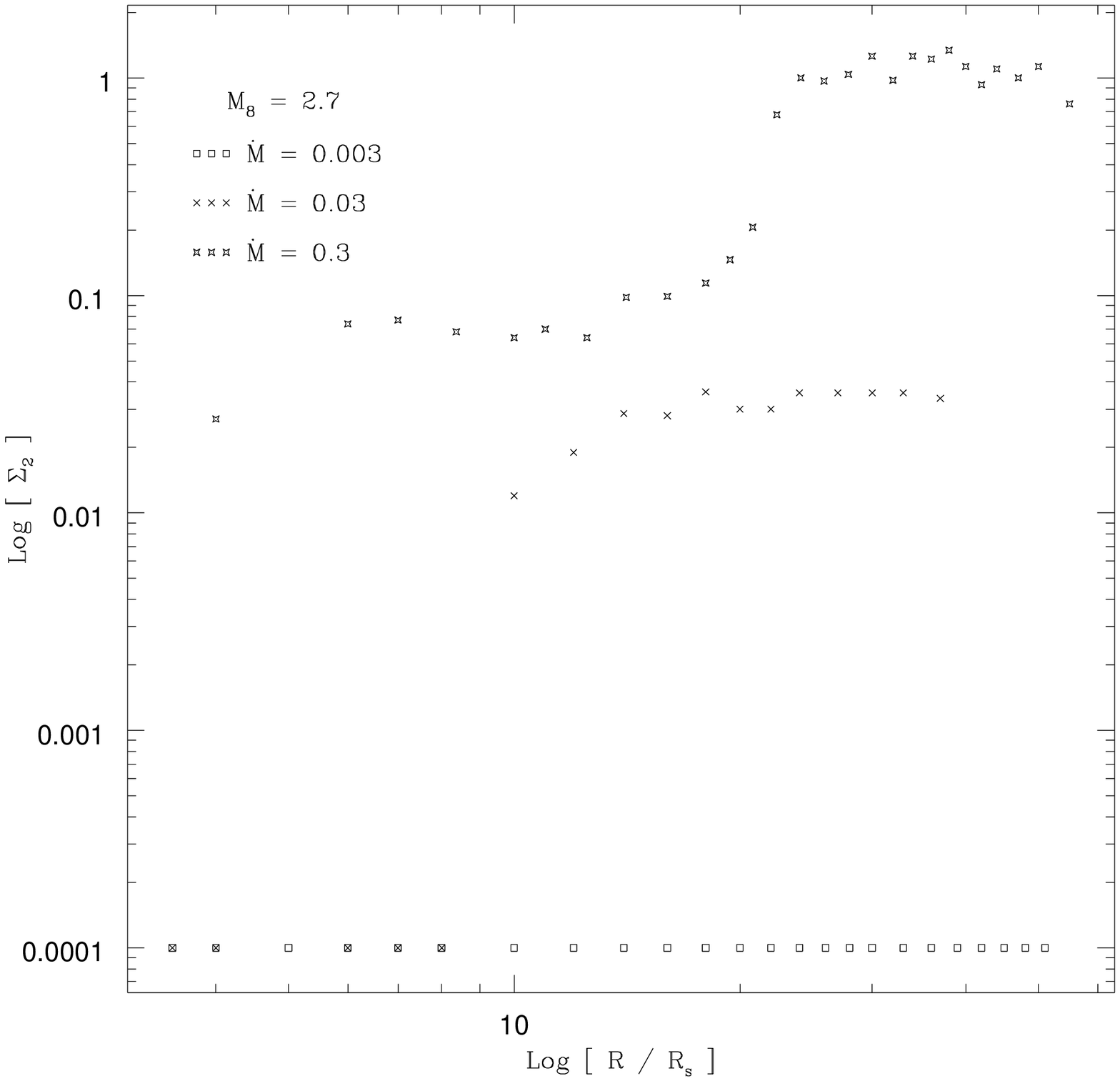}}
\caption{\label{fig: rsig mdot} }
\end{figure}

\bigskip
Figure \ref{fig: rsig mdot}.  The mass column of the thermal runaway 
($\Sigma_2$) for fixed $m_8$ and variable $\dot m$.
The outer boundary
of the integration is $\Sigma = 10^{-4} \mbox{g ${cm^{-2}}$}$. 

\begin{figure}
\centerline{\plotone{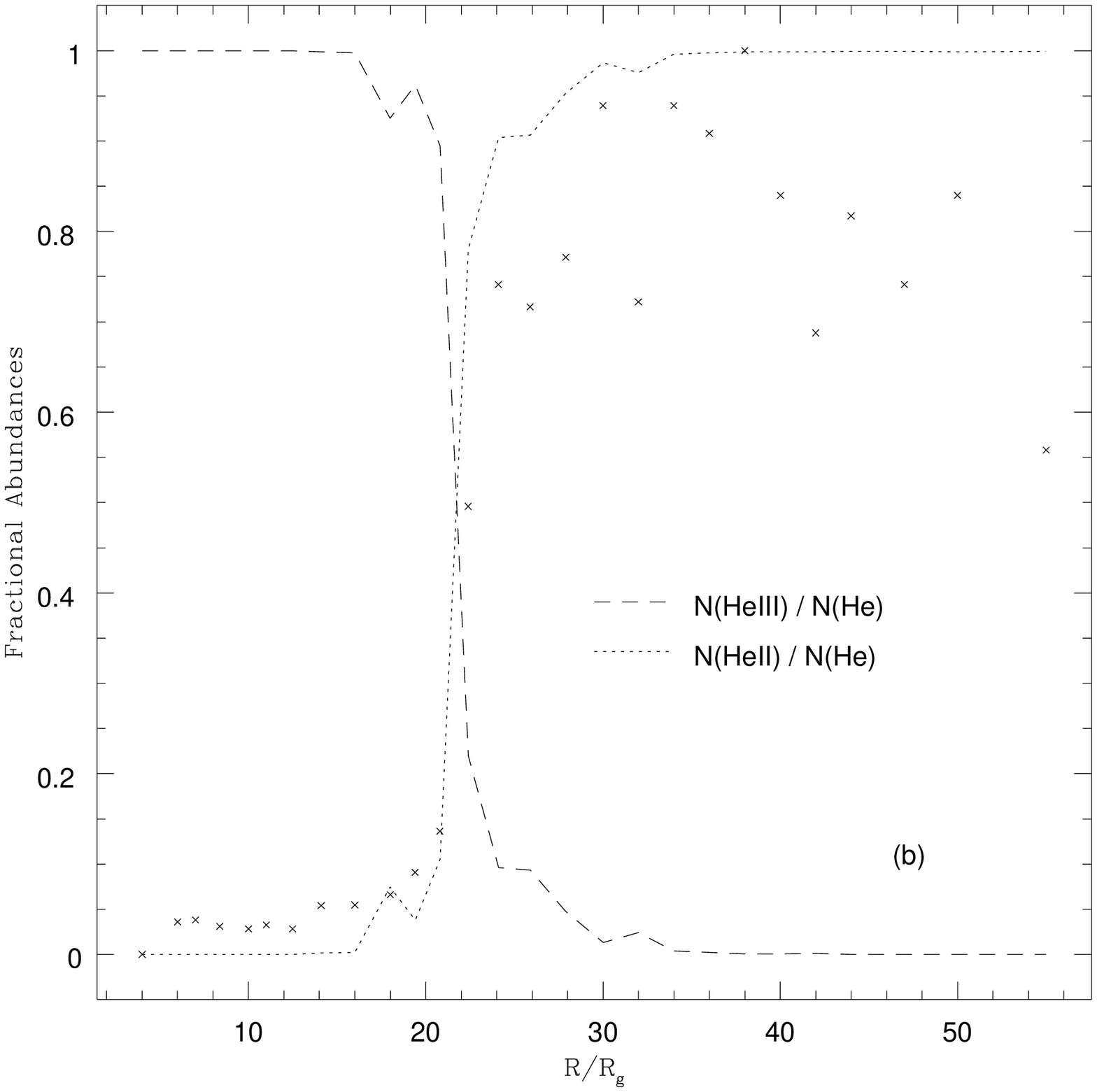}}
\caption{\label{fig: heIII abundance}}
\end{figure}

\bigskip
Figure \ref{fig: heIII abundance}.  The fractional abundances of HeII and HeIII 
at
$\Sigma_2$
for $\dot m = 0.3$ and $m_8=2.7$.  The points are $\Sigma_2$ for this model.

\begin{figure}
\centerline{\plotone{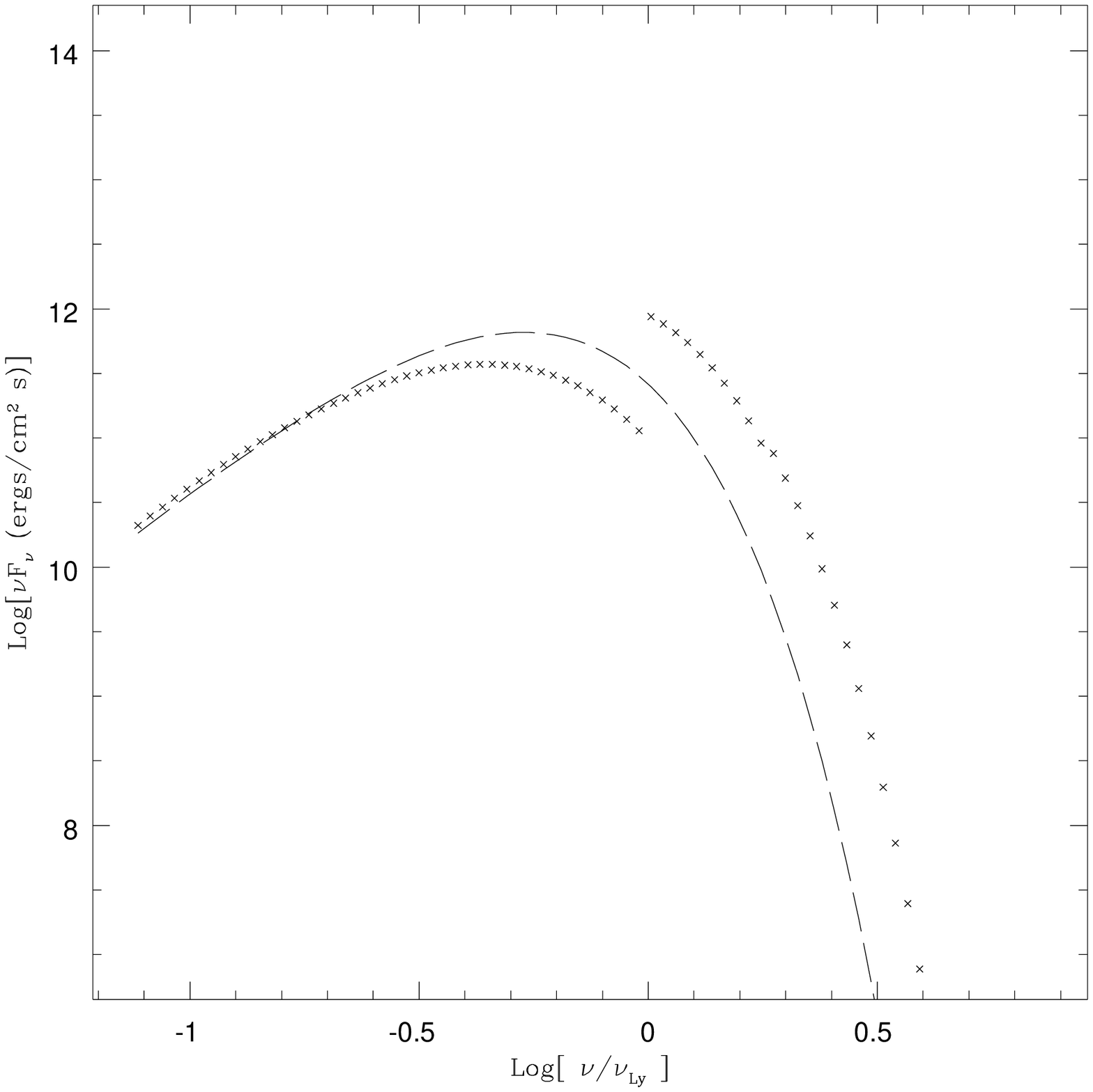}}
\caption{\label{fig: localspec2} }
\end{figure}

\bigskip
Figure \ref{fig: localspec2}. The spectrum at $r=14.0$ for $\dot m=0.03$
and $m_8=2.7$.  The crosses are the results of the simulation and the dashed
line represents a black body spectrum at the effective temperature of the 
disk.

\begin{figure}
\centerline{\plotone{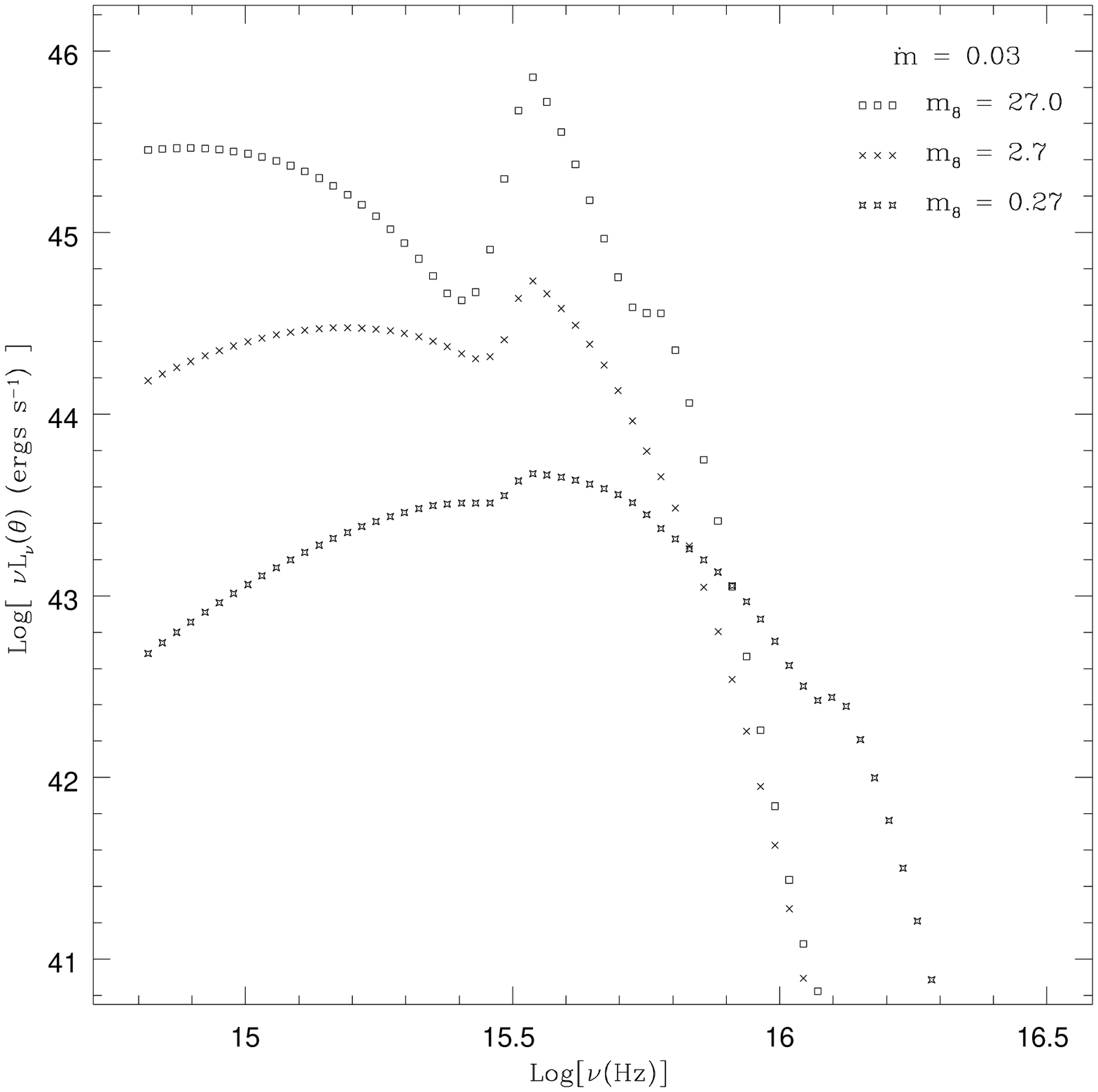}}
\caption{\label{fig: mass spectra} }
\end{figure}

\bigskip
Figure \ref{fig: mass spectra}. The face-on disk spectrum for a fixed 
mass accretion rate and varying central mass.

\begin{figure}
\centerline{\plotone{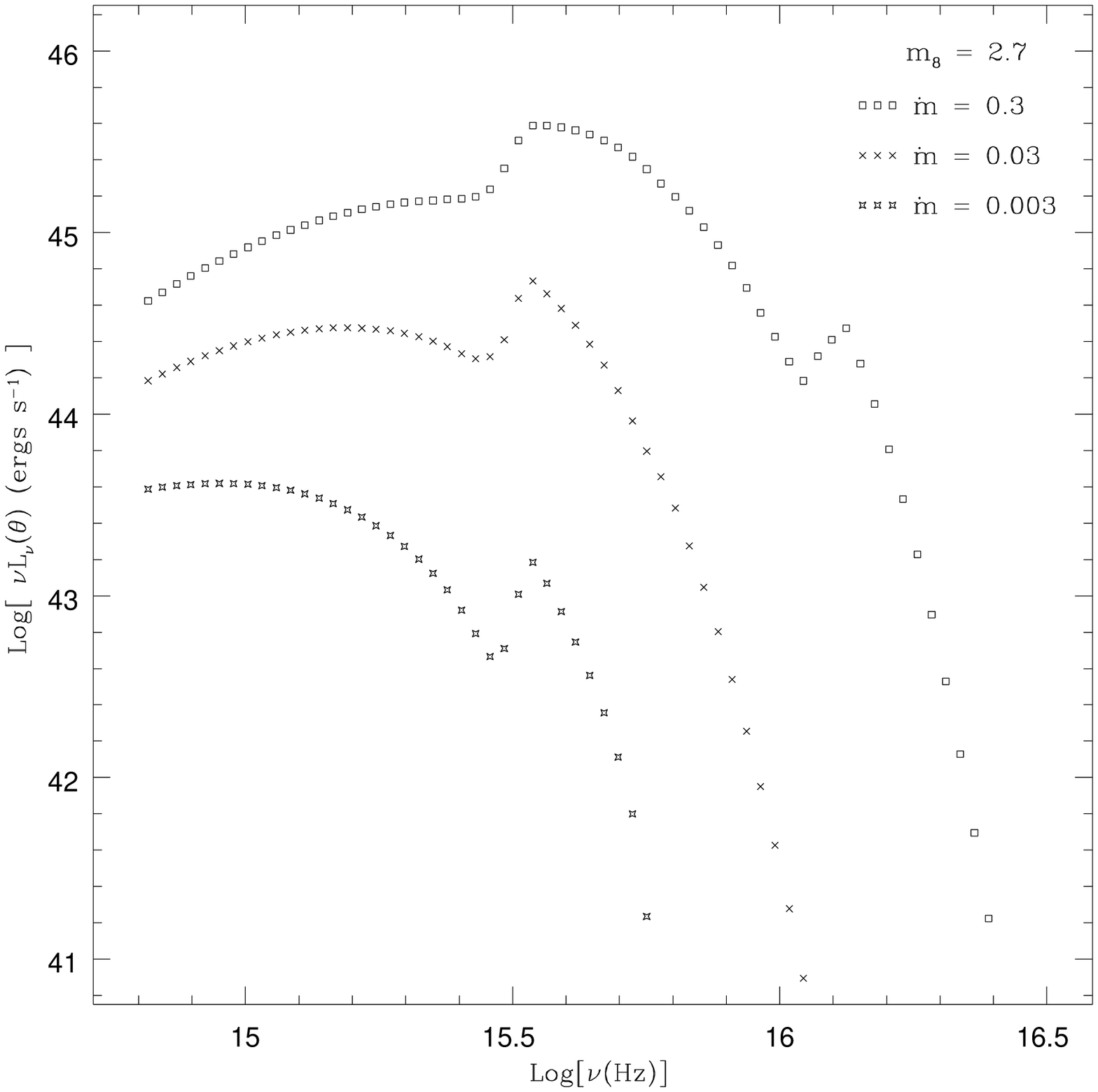}}
\caption{\label{fig: mdot spectra} }
\end{figure}

\bigskip
Figure \ref{fig: mdot spectra}.  The face-on disk spectrum for a fixed
central mass and varying mass accretion rate.

\begin{figure}
\centerline{\plotone{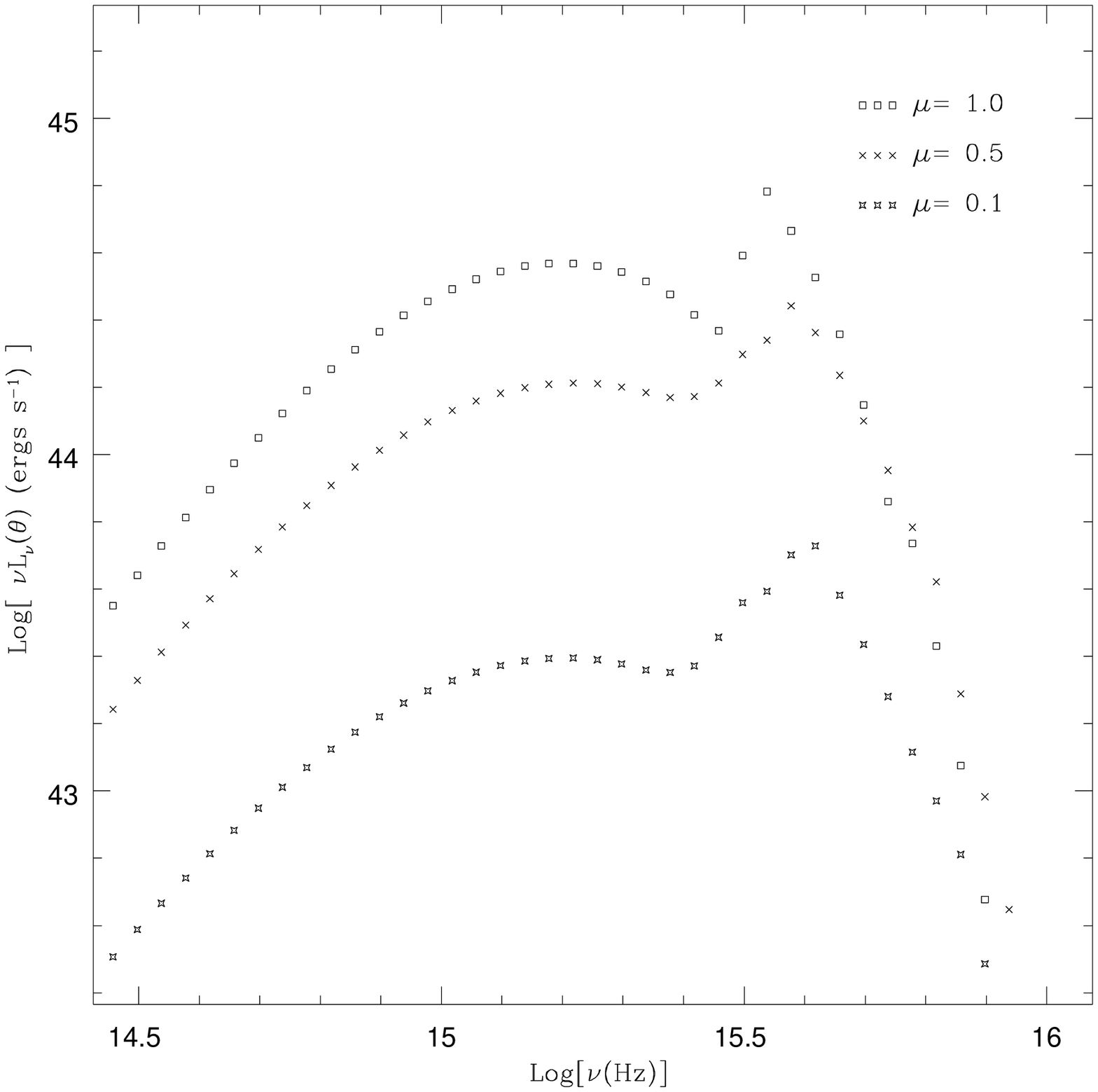}}
\caption{\label{fig: limb disk} }
\end{figure}

\bigskip
Figure \ref{fig: limb disk}. The disk spectrum as a function
of inclination, $\mu = \cos\theta$.

\begin{figure}
\centerline{\plotone{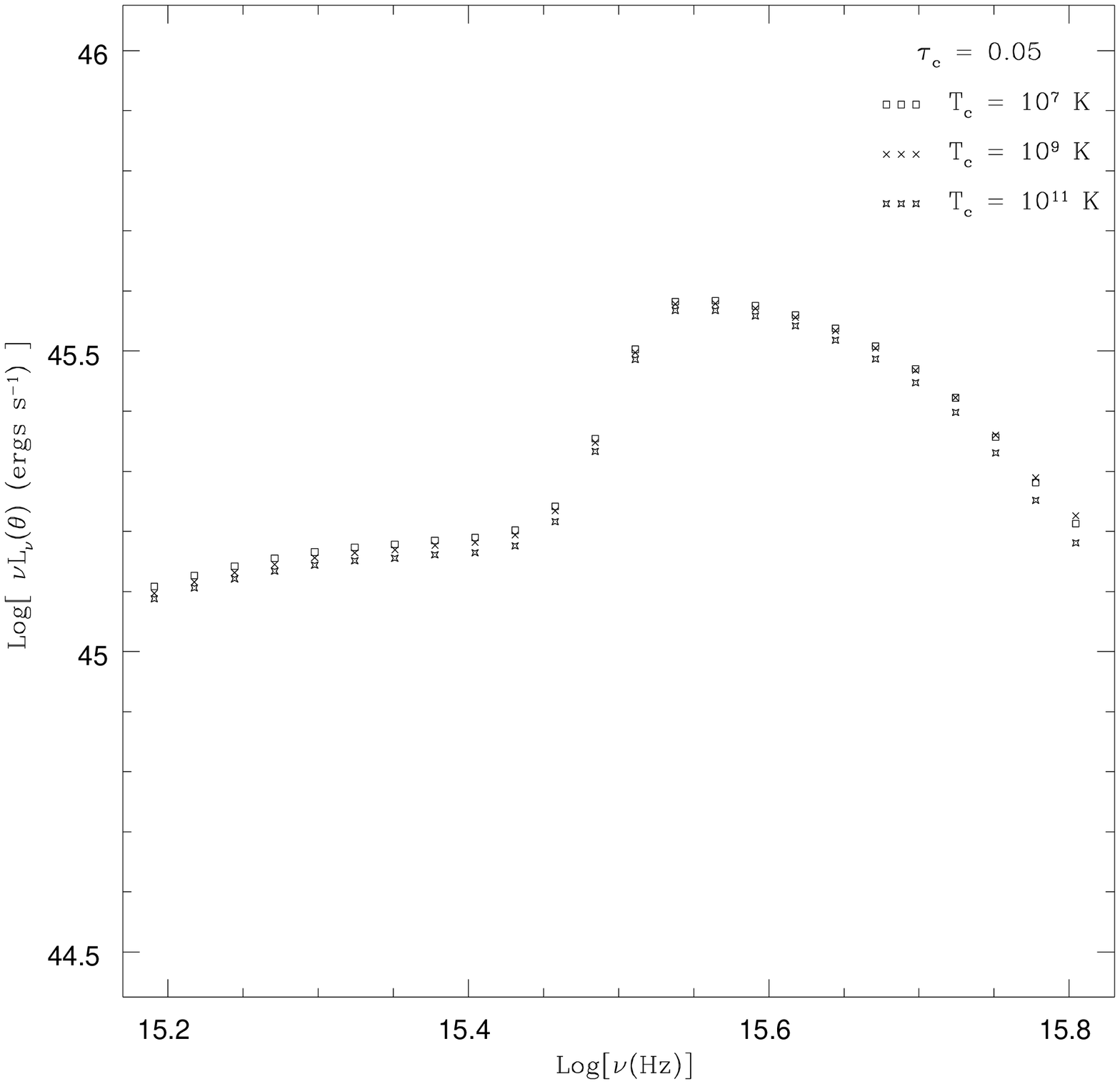}}
\caption{\label{fig: Compton t} }
\end{figure}

\bigskip
Figure \ref{fig: Compton t}.  The Comptonized UV spectrum of a face-on
disk for a fixed $\tau_c$
and varying $T_c$.  The disk parameters are $\dot m = 0.03$ and $m_8 = 2.7$.

\begin{figure}
\centerline{\plotone{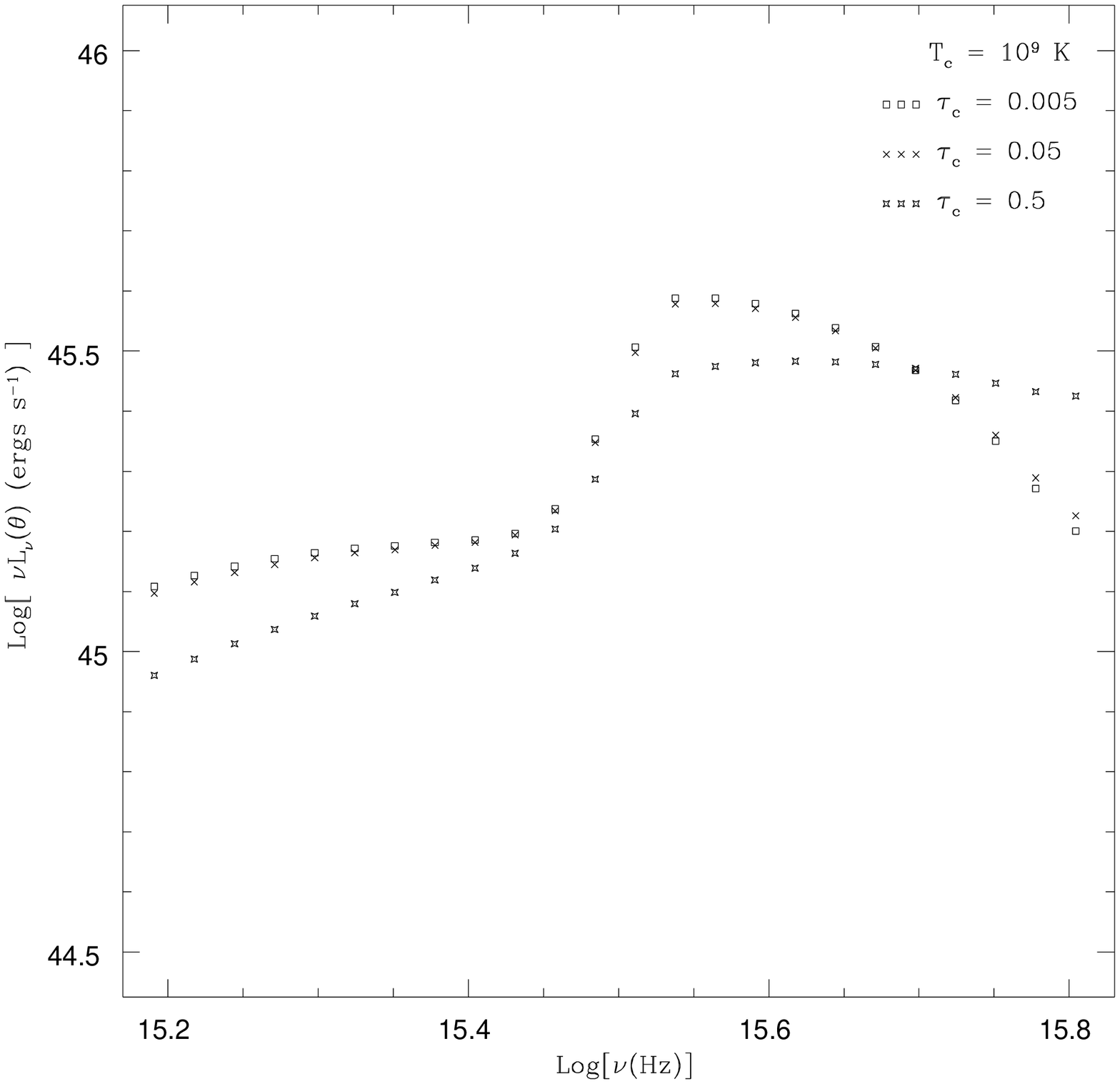}}
\caption{\label{fig: Compton tau} }
\end{figure}

\bigskip
Figure \ref{fig: Compton tau}.  The Comptonized UV spectrum of a face-on disk
for a fixed $T_c$
and varying $\tau_c$.  The disk parameters are $\dot m = 0.03$ and $m_8 = 2.7$.

\bigskip
Table \ref{table: uv fraction}.  The fraction of the incident X-ray flux
which is reprocessed into thermal UV emission ($f_{UV}$) and the continuum
spectral index ($\alpha_{15}$) for different models.

\vfil\eject
\begin{table}
\begin{tabular}{||c|c|c|c|c||} \hline
$\dot m$  & $m_8$   & $L_T$     & $f_{UV}$ & $\alpha_{15}$ \\ \hline
0.003     & 2.7     & $10^{44}$ & 1.0      &    -1.2       \\ 
0.03      & 0.27    & $10^{44}$ & 0.96     &     0.8       \\
0.03      & 2.7     & $10^{45}$ & 0.91     &    -0.2       \\
0.03      & 27.0    & $10^{46}$ & 0.85     &    -1.6       \\
0.3       & 2.7     & $10^{46}$ & 0.62     &     0.3       \\ \hline
\end{tabular}
\caption{\label{table: uv fraction}}
\end{table}

\end{document}